\documentclass{JHEP3}
\usepackage{amsmath}

\usepackage{epsfig}
\usepackage{amssymb}
\usepackage{graphics}
\usepackage{amsthm}
\usepackage{graphicx}

\newcommand{\bea}{\begin{eqnarray}}
\newcommand{\eea}{\end{eqnarray}}
\newcommand{\bean}{\begin{eqnarray*}}
\newcommand{\eean}{\end{eqnarray*}}
\newcommand{\nn}{\nonumber \\}

\def\W #1{\widetilde{#1}}
\def\WH #1{\widehat{#1}}

\def\eref#1{(\ref{#1})}

\def\d{{\rm d}}

\def\a{{\alpha}}

\def\d{\partial}

\def\eps{\epsilon}

\def\Label#1{\label{#1}%
  \smash{\hbox to0pt{\raise1ex\hbox{\tiny[#1]}\hss}}}

\preprint{USTC-ICTS/PCFT-20-43 }

\title{Reduction of one-loop integrals with higher poles by unitarity cut method}
\author{ Bo Feng$^{abc}$, Hongbin Wang$^{a}$
\footnote{Emails:  fengbo@zju.edu.cn, 21836003@zju.edu.cn . The corresponding author is
Hongbin Wang.} \\
{$^a$\small Zhejiang Institute of Modern Physics, Zhejiang University, Hangzhou, 310027, P. R. China \\
$^b$ Center of Mathematical Science, Zhejiang University, Hangzhou, 310027, P. R. China\\
$^c$ Peng Huanwu Center for Fundamental Theory, Hefei, Anhui 230026, China}}
\date{\today}
\abstract{ Unitarity cut method has been proved to be very useful in the computation of one-loop integrals. In this paper, we generalize the method to the situation where the powers of propagators
in the denominator are larger than one in general. We show how to use the trick of differentiation over
masses to translate the problem to the integrals where  all powers are just one.
Then by using the unitarity cut method, we can find the wanted reduction coefficients of all basis except the tadpole. Using this method, we calculate the reduction of scalar bubble, scalar triangle, scalar box and scalar pentagon with general power of propagators.
}


\newpage
\begin{document}

%%%%%%%%%%%%%%%%%%
\section{Motivation}
%%%%%%%%%%%%%%%%%%%%

Unitarity cut method \cite{Bern:1994zx,Bern:1994cg,Britto:2004nc} has been proved to be very efficient method to calculate one-loop amplitudes. The power of this method is that using the
holomorphic anomaly \cite{Cachazo:2004zb}, one can analytically carry out the reduced phase space integration with
double cuts, thus to extract the wanted reduction coefficients of master basis. The phase space integration method has been systematically developed for pure $4D$-dimension in \cite{Britto:2005ha,Britto:2006sj}  and for general $(4-2\eps)$-dimension in \cite{Anastasiou:2006jv,Anastasiou:2006gt,Britto:2006fc}. With these results, pure analytic
algebraic expressions for reduction coefficients have been given in \cite{Britto:2006fc,Britto:2007tt,Britto:2008vq} and their properties have been studied in \cite{Britto:2008sw,Feng:2013sa}.

All above works have been assumed that the power of propagators in the denominator of one-loop integrals is just one. This assumption is harmless for general situations, but higher power of propagators does appear in some situation\footnote{Some works for the higher power of propagators can be found in \cite{Sogaard:2014ila,Zhang:2011ns,Abreu:2017idw}}, for example, in higher loops or in the middle steps if one use the IBP method to do the reduction. Thus it is naturally to ask if the unitarity cut method can be
applied to these more general situations.

In this paper, we consider the reduction of one-loop integrals with general pole structures, i.e.,
propagators could have general power. The general integration will be given by
\bea {\cal M}[\ell] \equiv \int {d^D\ell \over (2\pi)^{D/2}} { {\cal N}[\ell]\over \prod_{j=1}^n ((\ell-K_j)^2- m_j^2+i\eps)^{a_j}},~~a_i\geq 1~~~~\label{gen-exp} \eea
where ${\cal N}[\ell]$ is an arbitrary polynomial function of $\ell$.
According to the PV-reduction method \cite{Passarino:1978jh}, we can decompose
\bea {\cal M}[\ell] =\sum_t c_t {\cal I}_t[\ell]~~~~\label{gen-im-0} \eea
where $c_t$'s are the rational functions  and ${\cal I}_t$ are scalar basis of tadpole, bubble,
triangle, box and pentagon.
Let us consider the $c_i$'s first. The data entering the integral  \eref{gen-exp} are the external momenta, the masses and the polarization vectors (the coupling constants are overall factor, so can be dropped in the discussion). The $c_i$ should be the rational functions of these data with Lorentz invariant contractions.

The idea of unitarity cut method is to compare the imaginary part of both sides of \eref{gen-im-0}. Since $c_t$'s are rational functions without imaginary contributions, we have
\bea {\rm Im}({\cal M}[\ell]) =\sum_t c_t {\rm Im}({\cal I}_t[\ell])~~~~\label{gen-im-1} \eea
Different master integrals have different analytic structures for the imaginary part (we will call them as the "signature"), thus if we can analytically computer the left hand side, we can do the spliting according to the analytic signature of each master integral and find expansion coefficients $c_t$ at the right hand side. With this thought, the key of reduction by unitarity cut method is to compute the imaginary part of ${\cal M}[\ell]$. When all $a_j=1$, the computation of left hand side \eref{gen-im-1} is transformed to
the reduced phase space integration
with double cuts, which we know how to do it as reviewed in the first paragraph of this section. For general $a_i$, we use following trick to solve the problem.
Noticing that\footnote{The same trick has also been used to get the homogenous solution of differential equations by maximum cut in \cite{Primo:2016ebd}.}
\bea & &  \int {d^D\ell \over (2\pi)^{D/2}} { {\cal N}[\ell]\over \prod_{j=1}^n ((\ell-K_j)^2- m_j^2+i\eps)^{a_i}}\nn
& = & \left\{\prod_{j=1}^n {1\over (a_j-1)!}{d^{a_j-1}\over d \eta_j^{a_j-1}}\int {d^D\ell \over (2\pi)^{D/2}} { {\cal N}[\ell]\over \prod_{j=1}^n ((\ell-K_j)^2- m_j^2-\eta_j+ i\eps)}\right\}|_{\eta_j\to 0}
~~~~\label{gen-im-2} \eea
the computation of the imaginary part of an one-loop integral is transformed to the reduced phase space integration
with double cut for the  case with all $a_i=1$. More explicitly, let us separate both sides of  \eref{gen-im-2} to the real and imaginary part, we have
\bea Re[L]+i Im[L]=\left\{\prod_{j=1}^n {1\over (a_j-1)!}{d^{a_j-1}\over d \eta_j^{a_j-1}} ( Re[R]+i Im[R])\right\}|_{\eta_j\to 0}~~~~\label{gen-im-3} \eea
Since the $\eta_i$'s are defined to take real values, we have
\bea  Re[L]+i Im[L]=\left\{\prod_{j=1}^n {1\over (a_j-1)!}{d^{a_j-1}\over d \eta_j^{a_j-1}}  Re[R]\right\}|_{\eta_j\to 0}+i \left\{\prod_{j=1}^n {1\over (a_j-1)!}{d^{a_j-1}\over d \eta_j^{a_j-1}} Im[R]\right\}|_{\eta_j\to 0}~~~~\label{gen-im-4} \eea
thus we get
\bea Im[L]=\left\{\prod_{j=1}^n {1\over (a_j-1)!}{d^{a_j-1}\over d \eta_j^{a_j-1}} Im[R]\right\}|_{\eta_j\to 0}~~~~\label{gen-im-5} \eea
For general ${\cal N}[\ell]$ in \eref{gen-im-2}, we know the expansion
\bea Im[R]=\sum_t  c_t Im({\cal I}_t[\ell])\eea
and the action of ${d\over d\eta}$ will act on both $c_t$ and $ Im({\cal I}_t[\ell])$.
Since the function $c_t$'s have been given in \cite{Britto:2006fc,Britto:2007tt,Britto:2008vq}, the unknown piece is the action of ${d\over d\eta}$ on $ Im({\cal I}_t[\ell])$ and its expansion. In another words, we just need to consider the reduction of general power with ${\cal N}[\ell]=1$ in  \eref{gen-exp} for $n\leq 5$.

The plan of the paper is following. In the section two, we consider the reduction of bubbles with higher poles. We establish the general recurrence relation and check our results with some examples. Same method has been applied to triangles, boxes and pentagons in the section three, four and five. A brief conclusion is given in the section six.

%%%%%%%%%%%%%%%%%%%%%%%%%%%%%%%%%%
\section{Bubble}
\label{sec:intro}
%%%%%%%%%%%%%%%%%%%%%%%%%%%%%%%%%%%

For bubble topology, let us define
\bea I_2(a,b)[K;M_1, M_2]\equiv \int \frac{d^{4-2\epsilon}p}{(2\pi)^{4-2\epsilon}}\frac{1}{(p^2-M_1^2)^a((p-K)^2-M_2^2)^b}
~~~\label{Bu-gen-1-1} \eea
to be the general scalar bubble integral with higher power of propagators. The master integral of bubble is the case $n=m=1$ and for this special case, sometimes we just write it as
${\cal I}_2[K;M_1, M_2]$ or just ${\cal I}_2$ for simplicity. The reduction of $I_2(a,b)[K;M_1, M_2]$ will be the following expansion
\bea I_2(a,b)[K;M_1, M_2] & = & c_{2\to 2}(a,b) {\cal I}_{2}[K;M_1, M_2]+
\sum_{i=1}^2 c_{2\to 1;i}(a,b) {\cal I}_{1;i}[M_i] ~~~\label{Bu-gen-1-1-1}  \eea
where $c_{2\to 1;i}$ means that when reducing the bubble to the tadpole, the $i$-th propagator has been kept. The tadpole ${\cal I}_{1;i}[M_i]$ should be written in the standard form with the proper momentum shifting, i.e.,
\bea {\cal I}_{1;i}[M_i]\equiv \int \frac{d^{4-2\epsilon}p}{(2\pi)^{4-2\epsilon}}\frac{1}{(p^2-M_i^2)}
~~~\label{Bu-gen-1-1-2} \eea
Expansion coefficients in \eref{Bu-gen-1-1-1} can be found by various methods, for example, the IBP method. However, in this paper, we will try to use the unitarity cut method to find expansion coefficients. When we use the unitarity cut method, the tadpole can not be detected, thus tadpole coefficients can not be found by this way. Although we will not consider the tadpole coefficients in this paper, we want to point out that some efforts have been done to fill the gap by using the single cut \cite{Britto:2009wz,Britto:2010um,Britto:2012mm}.

The unitarity cut of $I_2(1,1)$ is given by (see references \cite{Anastasiou:2006jv,Britto:2006fc,Anastasiou:2006gt})\footnote{In the reference \cite{Britto:2006fc} when we do the general one-loop unitarity cut phase space integration, an overall factor $\left( {\Delta[K,M_1,M_2]\over K^2}\right)^{-\eps}$ has been neglected in later computation (see the equation after eq.(9)). This is fine for the work in \cite{Britto:2006fc}, but since it depends on the masses, it is crucial for current computation and we must include it back.}
\bea {\cal C}[ {\cal I}_2]=(K^2)^{-1+\eps}\Delta^{\frac{1}{2}-\epsilon} \int_0^1 \mathrm{d}uu^{-1-\epsilon}\sqrt{1-u}~~~\label{Bu-gen-1-2}\eea
where
\bea \Delta[K;M_1, M_2]&= & {(K^2)}^2+({M_1}^2)^2+({M_2}^2)^2-2{M_1}^{2}{M_2}^2-2K^{2}{M_1}^2-2K^{2}M_{2}^2\nn
& = & -4 M_1^2 M_2^2 \left| \begin{array}{cc} 1 &
-{K^2-M_1^2-M_2^2\over 2M_1 M_2}\\ -{K^2-M_1^2-M_2^2\over 2M_1 M_2}
& 1 \end{array}\right|~.~~\label{Bu-gen-1-3}\eea
which is the Landau surface of bubble of the first type of singularities.
For the later convenience, let us define
\bea Bub^{(n)}=\frac{1}{2(n-\epsilon)}\int_0^1du\frac{u^{n-\epsilon}}{\sqrt{1-u}}=
\int_0^1 \mathrm{d}uu^{-1-\epsilon} u^n \sqrt{1-u}~.~~\label{Bu-gen-1-2-1}\eea
Thus by comparing \eref{Bu-gen-1-2-1} with \eref{Bu-gen-1-2} we see that ${\cal C}[{\cal I}_2]=(K^2)^{-1+\eps}\Delta^{\frac{1}{2}-\epsilon} Bub^{(0)}$. The $Bub^{(n)}$ is well defined for
$n\geq 0$ and it is easy to derive a recursion relation by integration-by-part
\bea Bub^{(n)}= { (n-1-\eps)\over (n+{1\over
2}-\eps)}Bub^{(n-1)}.~~~\label{I-2m-recu-1}\eea
Solving it  we get
\bea Bub^{(n)}  =  F^{(n)}
Bub^{(0)},~~~~~F^{(n)} & = & {\Gamma(3/2-\eps) \Gamma(n-\eps)\over
\Gamma(-\eps)\Gamma(n+3/2-\eps)}~.~~~~\label{I-2m-recu-3} \eea
We find that when doing the reduction for triangles and boxes, we will meet the form $Bub^{(n)}$ with the negative integer $n$. For this case, we can use \eref{I-2m-recu-1} to analytically continue from positive $n$ to negative $n$.
%But now the integration is not well defined because the divergence around $u=0$. For this case, we can analytically continue from larger negative $\eps$ to $\eps\to 0^-$ and use \eref{I-2m-recu-1} under this understanding.
For example, using $n=0$ in \eref{I-2m-recu-1}, we get
\bea Bub^{(-1)}= {(1-2\eps)\over 2(-1-\eps)}  Bub^{(0)}~~~~\label{I-2m-recu-4} \eea
%
%One can check \eref{I-2m-recu-4} as following
%
%\bea Bub^{(0)}-2\epsilon Bub^{(0)} & = & %\int_0^1`duu^{-1-\epsilon}\sqrt{1-u}+\int_0^1du\frac{u^{-\epsilon}}{\sqrt{1-u}}\nn
%
%& = & \int_0^1du^{-1-\epsilon}\frac{1}{\sqrt{1-u}}= 2(-1-\eps) Bub^{(-1)}
%~~~~\label{I-2m-recu-5} \eea
%

Having above preparation, now we consider the reduction of $I_{2}(a,b)$.
According to our general idea in \eref{gen-im-5}, we should calculate\footnote{From the discussion of \eref{gen-im-2} one can see that the role of $\eta$ is identical with $m^2$. For scalar basis,
$m$ does not appear in other places, so we can take the derivative of $m^2$ instead of $\eta$ without making
any mistake. } $\left(\frac{\partial}{\partial M_1^2}\right)^{a-1}\left(\frac{\partial}{\partial M_2^2}\right)^{b-1} \Delta^{\frac{1}{2}-\epsilon}$. To get a better idea, let us start with
$\left(\frac{\partial}{\partial M_1^2}\right)^{n} \Delta^{\frac{1}{2}-\epsilon}$.
Before doing this, we rewrite $\Delta$ as
\bea
\Delta
&=&(M_1^2-M_2^2-K^2-\sqrt{2K^2M_2^2})(M_1^2-M_2^2-K^2+\sqrt{2K^2M_2^2})~~~\label{Bu-gen-1-4}
\eea
where each factor is linear in $M_1^2$. It is easy to get the $n$-th derivative of $M_1^2$
is given by
\bea
\left(\frac{\partial}{\partial M_1^2}\right)^n \Delta^{\frac{1}{2}-\epsilon}
& = & \sum_{\lambda=0}^n C_n^{\lambda}\left(\frac{\partial}{\partial M_1^2}\right)^{n-\lambda}\left\{(M_1^2-M_2^2-K^2-\sqrt{2K^2M_2^2})^{\frac{1}{2}-\epsilon}\right\}\nn
& &\times \left(\frac{\partial}{\partial M_1^2}\right)^{\lambda}\left\{(M_1^2-M_2^2-K^2+\sqrt{2K^2M_2^2})^{\frac{1}{2}-\epsilon}\right\}
~~~\label{Bu-gen-2-1}\eea
Using $\Gamma(x+1)= x\Gamma(x)$, we have
\bea {d^n (x-a)^b\over dx^n}= b(b-1)..(b-n+1) (x-a)^{b-n}= {\Gamma(b+1)\over \Gamma(b+1-n)}(x-a)^{b-n}~~~\label{Bu-gen-2-2} \eea
thus  \eref{Bu-gen-2-1} is
\bea
\left(\frac{\partial}{\partial M_1^2}\right)^n \Delta^{\frac{1}{2}-\epsilon}
& = & \Delta^{{1\over 2}-\eps} \left\{\sum_{\lambda=0}^n C_n^{\lambda} {\Gamma({3\over 2}-\eps)\over \Gamma({3\over 2}-\eps-(n-\lambda))}\Delta_{M_1,-}^{-(n-\lambda)}
{\Gamma({3\over 2}-\eps)\over \Gamma({3\over 2}-\eps-\lambda)}\Delta_{M_1,+}^{-\lambda}\right\}~~~\label{Bu-gen-2-4}\eea
where we have defined
 \bea\Delta_{M_1,\pm}= M_1^2-M_2^2-K^2\pm \sqrt{2K^2M_2^2}~~~\label{Bu-gen-2-5}\eea

Using above result, we can find the reduction coefficient of $I_2(n+1,1)$ by \eref{gen-im-5}
since
\bea {\cal C}[I_2(n+1,1)] & = &\frac{1}{n!}\left(\frac{\partial}{\partial M_1^2}\right)^n{\cal C}[{\cal I}_2]= c_{2\to 2}(n+1,1) {\cal C}[{\cal I}_2] ~~~\label{Bu-gen-3-1}\eea
where
\bea c_{2\to 2}(n+1,1)[K;M_1,M_2]={1\over n!} \left\{\sum_{\lambda=0}^n C_n^{\lambda} {\Gamma({3\over 2}-\eps)\over \Gamma({3\over 2}-\eps-(n-\lambda))}\Delta_{M_1,-}^{-(n-\lambda)}
{\Gamma({3\over 2}-\eps)\over \Gamma({3\over 2}-\eps-\lambda)}\Delta_{M_1,+}^{-\lambda}\right\}~~~\label{Bu-gen-3-2}\eea
by the result \eref{Bu-gen-2-4}.

For the general bubble coefficients of the reduction of $ I_{2}(n+1,m+1)$, using the relation
\bea
{\cal C}[I_2(n+1,m+1)] & = &
\frac{1}{m!n!}\left(\frac{\partial}{\partial M_2^2}\right)^{m}\left(\frac{\partial}{\partial M_1^2}\right)^n{\cal C}[I_2(1,1)] ~~~\label{Bu-gen-4-1} \eea
we get reduction coefficient as
\bea
c_{2\to 2}(n+1,m+1)[K;M_1, M_2] &= & \frac{1}{m!n!\Delta^{\frac{1}{2}-\epsilon}}\left(\frac{\partial}{\partial M_2^2}\right)^{m}\left(\frac{\partial}{\partial M_1^2}\right)^n\Delta^{\frac{1}{2}-\epsilon}~~~\label{Bu-gen-4-2} \eea
The general analytic expression of $c_{2\to 2}(n+1,m+1)$ will be complicated to write down.

Now we check the result \eref{Bu-gen-4-2}. The first check is by the symmetry. Noticing that when shifting $p\to p+K$, the \eref{Bu-gen-1-1} becomes
\bea \int \frac{d^{4-2\epsilon}p}{(2\pi)^{4-2\epsilon}}\frac{1}{(((p+K)^2-M_1^2)^a(p^2-M_2^2)^b}
~~~\label{Bu-gen-6-1} \eea
thus we have $I_2(a,b)[K;M_1,M_2]= I_2(b,a)[-K;M_2,M_1]$. When doing reduction at both sides, we must have the bubble coefficient to be same, i.e,
\bea c_{2\to 2}(n+1,m+1)[K;M_1,M_2]= c_{2\to 2}(m+1,n+1)[-K;M_2,M_1]~~~\label{Bu-gen-6-2} \eea
From the first line of \eref{Bu-gen-4-2}, the $\Delta$ is invariant under the replacement $[K;M_1,M_2]\to [-K;M_2,M_1]$, so it does satisfy the relation \eref{Bu-gen-6-2}.

The second check is following.  For one loop massless integral (example 5.2 of \cite{Smi:Eva})
\bea F_{5.2}(a_1, a_2)=\int {d^d k\over (k^2)^{a_1} [(q-k)^2]^{a_2}}~~~\label{Bu-gen-5-1} \eea
by the IBP method, one can derive the relation
\bea F(a_1, a_2>1)=-{1\over (a_2-1) q^2}[(d-2a_1-a_2+1) F(a_1, a_2-1)- (a_2-1) F(a_1-1, a_2)]~~~\label{Bu-gen-5-2} \eea
and when $a_2=1$ we have
\bea F(a_1, 1)= -{d-a_1-1\over (a_1-1)q^2} F(a_1-1,1)~.~~\label{Bu-gen-5-3}  \eea
To compare with our calculation, we should take $M_1, M_2\to 0$, $K\to q$ and $d\to 4-2\eps$ after all derivatives
related to $M_1^2, M_2^2$ having been done. We have used the Mathematica to check the massless case of our general result \eref{Bu-gen-4-2}. The third check is that we have used the LiteRed \cite{Lee:2013mka} to explicitly calculate some examples with nonzero $M_1, M_2$ and we do find the match.
%%%%%%%%%%%%%%%%%%
\subsection{Recurrence relation}
%%%%%%%%%%%%%%%%%%%

Like the IBP relation, the idea of \eref{gen-im-5} can be used to establish the recurrence relation for reduction coefficients. Let us start from the reduction of $I_2(a,1)$. Noticing that
\bea I_2(a,1) & = & \int \frac{d^{4-2\epsilon}p}{(2\pi)^{4-2\epsilon}}\frac{1}{(p^2-M_1^2)^a((p-K)^2-M_2^2)}\nn
& = &{1\over (a-1)} {d\over d (M_1^2)} {1\over (a-2)!} {d^{a-2}\over d (M_1^2)^{a-2}}\int \frac{d^{4-2\epsilon}p}{(2\pi)^{4-2\epsilon}}\frac{1}{(p^2-M_1^2)((p-K)^2-M_2^2)} \nn
&= & {1\over (a-1)} {d\over d (M_1^2)} I_2(a-1,1)~,~~\label{Bubble-rec-1-1}\eea
if we know
\bea I_2(a-1,1) = c_{2\to 2}(a-1,1)[K; M_1,M_2] I_2(1,1)+... ~~~\label{Bubble-rec-1-2}\eea
where $...$ is for the tadpole part,
we can write
\bea I_2(a,1) & = & \left({1\over (a-1)} {dc_{2\to 2}(a-1,1)\over d (M_1^2)}\right) I_2(1,1)+ {c_{2\to 2}(a-1,1)\over (a-1)} I_2(2,1)+...~~~~~\label{Bubble-rec-1-3}\eea
where for simplicity we have dropped  the dependence of $K,M$.
From \eref{Bubble-rec-1-3}, we can read out a recurrence relation for the  coefficient\footnote{
The relation \eref{Bubble-rec-1-4} holds for $a\geq 2$, but when $a=2$, the $c_{2\to 2}(1,1)= 1$ and the identity is trivial.}
\bea c_{2\to 2}(a,1)[K; M_1,M_2] & = & \left({1\over (a-1)} {dc_{2\to 2}(a-1,1)\over d (M_1^2)}\right)+{c_{2\to 2}(a-1,1)\over (a-1)} c_{2\to 2}(2,1)
~~~\label{Bubble-rec-1-4}\eea
Relation \eref{Bubble-rec-1-4} tell us that staring from $c_{2\to 2}(2,1)$, we can write down all $c_{2\to 2}(a,1)$.

The same idea can be used to write down the recurrence relation of general $c_{2\to 2}(a,b)$. Noticing that
\bea I_2(a,b) & = &  ={1\over (b-1)!} {d^{b-1}\over d (M_2^2)^{b-1}} I_{2}(a,1) = {1\over (b-1)!} {d^{b-1}\over d (M_2^2)^{b-1}} \left\{ c_{2\to 2}(a,1) I_{2}(1,1)+...\right\}\nn
& = & {1\over (b-1)!} \sum_{t=0}^{b-1} C_{b-1}^t  {d^t c_{2\to 2}(a,1)\over d (M_2^2)^{t}}
(b-1-t)! I_2(1,b-t)+...~~\label{Bubble-rec-2-1}\eea
we get immediately the recurrence relation
\bea c_{2\to 2}(a,b) & = & {1\over (b-1)!} \sum_{t=0}^{b-1} C_{b-1}^t (b-1-t)!c_{2\to 2}(1,b-t) {d^t c_{2\to 2}(a,1)\over d (M_2^2)^{t}}~~\label{Bubble-rec-2-2}\eea
where $c_{2\to 2}(1,m)$ can be obtained from $c_{2\to 2}(m,1)$ by proper replacement as given in \eref{Bu-gen-6-2}. Result \eref{Bubble-rec-2-2} shows that knowing  the coefficient $c_{2\to 2}(2,1)$ is enough for general reduction coefficients. The similar idea can be used to the reduction for triangle, box and pentagon as will be shown later.

%%%%%%%%%%%%%%%%%%%%
\section{Triangle}
%%%%%%%%%%%%%%%%%%%%

For triangle topology, let us define\footnote{Although by momentum conservation, we will
have $K_1+K_2+K_3=0$. Keeping them all in the parameter will make the symmetry more transparent. }
\bea I_3(n_1,n_2,n_3)[K_1,K_2,K_3;M_1, M_2, m_1]=\int\frac{d^{4-2\epsilon}p}{(2\pi)^{4-2\epsilon}}\frac{1}{(p^2-M_1^2)^{n_1}
((p-K_1)^2-M_2^2)^{n_2}((p+K_3)^2-m_1^2)^{n_3}}~~~~~\label{Tri-gen-1-1} \eea
For the special case, i.e., $n_1=n_2=n_3=1$, we get the familiar scalar triangle basis, which for simplicity, we will denote as ${\cal I}_3[K_1,K_2,K_3;M_1, M_2, m_1]$. The unitarity cut of ${\cal I}_3$ with the cut momentum $K_1$ has been given by \cite{Britto:2006fc}
\bea {\cal C}(I_3)=-(\frac{4K_1^2}{\Delta[K_1,M_1,M_2]})^{\epsilon}\frac{1}{\sqrt{\Delta_{3;m=0}}}
Tri^{(0)}(Z)~~~\label{Tri-gen-2-1} \eea
with
\bea Tri^{(n)}(Z)=\int_0^1duu^{-1-\epsilon}u^nln\left(\frac{Z+\sqrt{1-u}}{Z-\sqrt{1-u}}\right)
=\frac{-Z}{n-\epsilon}\int_0^1du\frac{u^{n-\epsilon}}{\sqrt{1-u}(1-u-Z^2)},~~~n\geq 0~~~\label{Tri-gen-2-2} \eea
where $Z$ is given by
\bea Z=-\frac{(2K_1\cdot Q)K_1^2}{\sqrt{\Delta_{3;m=0}\Delta[K_1,M_1,M_2]}}~~~\label{Tri-gen-2-3} \eea
with\footnote{ The  $\Delta[K_1,M_1,M_2]$ reflects the singularity of bubble as  given in  \eref{Bu-gen-1-3},  while
$\Delta_{3;m=0}$ corresponds the pure second-type Landau singularity of triangle.}
\bea \Delta_{3;m=0} &= & 4((K_1\cdot K_3)^2-K_1^2K_3^2)=-4{\rm det}\left( \begin{array}{cc} K_1^2 & K_1\cdot K_3 \\ K_1\cdot K_3 & K_3^2\end{array} \right)~~~\label{Tri-gen-2-4} \\
\Delta[K_1,M_1,M_2] & = & (K_1^2)^2+(M_1^2)^2+(M_2^2)^2-2K_1^2M_1^2-2K_1^2M_2^2-2M_1^2M_2^2~~~\label{Tri-gen-2-5} \\
K_1\cdot Q &= & \frac{K_1^2+M_1^2-M_2^2}{K_1^2}(K_1\cdot K_3)+(K_3^2+M_1^2-m_1^2)~~~\label{Tri-gen-2-6} \eea
The $Tri^{(n)}(Z)$ has the recurrence relation \cite{Anastasiou:2006gt,Britto:2006fc}
%%%
\bea
Tri^{(n)}(Z)&=&-\frac{(Z^2-1)(n-1-\eps)}{n-\eps}Tri^{(n-1)}(Z)+\frac{2Z(n-1-\eps)}
{(n-\eps)}Bub^{(n-1)}
~~~\label{tri-rec}
\eea
%%%
which will be useful later.

For general $n_i$, the $I_3(n_1,n_2,n_3)$ will be reduced to the sum of one triangle $I_3$, three bubbles $I_{2;\bar{i}},i=1,2,3$ (where $\bar{i}$ means to set $n_i=0$ in \eref{Tri-gen-1-1} and other $n_j$'s one) and three tadpoles $I_{1;i},i=1,2,3$ (where $i$ means to set $n_i=1$ in \eref{Tri-gen-1-1} and other $n_j$'s zero). In other words, we will have the expansion
\bea I_{3}(n_1,n_2,n_3)= c_{3\to 3}(n_1, n_2, n_3) {\cal I}_{3} +
\sum_{i=1}^3 c_{3\to 2;\bar{i}}(n_1, n_2, n_3) {\cal I}_{2;\bar{i}}+ \sum_{i=1}^3 c_{3 \to 1;i}( n_1, n_2, n_3) {\cal I}_{1;i}~~~\label{Tri-gen-3-1} \eea
where for simplicity we have dropped the dependence over $K,M$  of coefficients $c$'s.
We want to emphasize that when writing down ${\cal I}_{2;\bar{i}}$, we need to do the proper momentum shifting to reach the standard form \eref{Bu-gen-1-1}. More explicitly, we will have
\bea {\cal I}_{2; \bar{3}} & = & \int\frac{d^{4-2\epsilon}p}{(2\pi)^{4-2\epsilon}}\frac{1}{(p^2-M_1^2)
((p-K_1)^2-M_2^2)} ~~~\label{Tri-gen-3-2} \eea
and
\bea {\cal I}_{2; \bar{1}} & = & \int\frac{d^{4-2\epsilon}p}{(2\pi)^{4-2\epsilon}}\frac{1}{
(p^2-M_2^2)((p-K_2)^2-m_1^2)}~~~\label{Tri-gen-3-3} \eea
by using the shifting $p\to p+K_1$, and finally
\bea {\cal I}_{2; \bar{2}} & = &\int\frac{d^{4-2\epsilon}p^4}{(2\pi)^{4-2\epsilon}}\frac{1}{((p-K_3)^2-M_1^2)
(p^2-m_1^2)}~~~\label{Tri-gen-3-4} \eea
by using the shifting $p\to p+K_1+K_2$. For later convenience, let us give a more general
bubbles coming from the triangle reduction \eref{Tri-gen-1-1}
% % % % % % %
\bea
&&I_{2;\bar 3}(n_1,n_2)[K_1;M_1, M_2]=\int\frac{d^{4-2\epsilon}p}{(2\pi)^{4-2\epsilon}}\frac{1}{(p^2-M_1^2)^{n_1}
	((p-K_1)^2-M_2^2)^{n_2}} \nn
&&I_{2;\bar 1}(n_1,n_2)[K_2; M_2, m_1]=\int\frac{d^{4-2\epsilon}p}{(2\pi)^{4-2\epsilon}}\frac{1}{(p^2-M_2^2)^{n_1}
	((p-K_2)^2-m_1^2)^{n_2}}\nn
&&I_{2;\bar 2}(n_2,n_1)[K_3;m_1, M_1]=\int\frac{d^{4-2\epsilon}p}{(2\pi)^{4-2\epsilon}}\frac{1}{((p-K_3)^2-M_1^2)^{n_1}
	(p^2-m_1^2)^{n_2}}~~~\label{Tri-gen-3-gen}
\eea

When we use the unitarity cut method to find the reduction coefficients, a given cut can only detect these coefficients of basis which have the corresponding cut. For each cut, it can detect the triangle coefficient and one bubble coefficient, while all tadpole coefficients can not be found by this way. For triangle, there are three different cuts, i.e, $K_1, K_2, K_3$.  the coefficient of bubble ${\cal I}_{2; \bar{3}}$ can only be detected by the cut with momentum $K_1$ and similarly bubble coefficients of ${\cal I}_{2; \bar{1}}$ and ${\cal I}_{2; \bar{2}}$ by the cut $K_2$ and $K_3$ respectively. Thus  using them all, we can get
reduction coefficients of triangle and bubbles. Furthermore, there are overlaps of detected reduction coefficients between different cuts. For the triangle case, all three cuts can detect the same triangle coefficient, so we can use the overlap as the cross check.

%%%%%%%%%%%%%%%%%%%%%
\subsection{ Recurrence relation}~~\label{Tri-recur}
%%%%%%%%%%%%%%%%%%%%%
Before going to explicit calculation, let us establish some recurrence relations like the one
\eref{Bubble-rec-1-4} and  \eref{Bubble-rec-2-2} in the previous section. To prepare for this task,
we discuss some symmetric properties of the integral \eref{Tri-gen-1-1} first. By shifting the momentum $p$, we can rewrite \eref{Tri-gen-1-1} to different forms, for example, by shifting
$p\to p+K_1$ we get
\bea I_3(n_1,n_2,n_3)[K_1,K_2,K_3;M_1,M_2,m_1] & = & \int\frac{d^{4-2\epsilon}p^4}{(2\pi)^{4-2\epsilon}}\frac{1}{((p+K_1)^2-M_1^2)^{n_1}
(p^2-M_2^2)^{n_2}((p-K_2)^2-m_1^2)^{n_3}}\nn
& = &  I_3(n_2, n_3, n_1)[K_2,K_3,K_1; M_2,m_1, M_1]~~~\label{Tri-gen-shift-1-1} \eea
and similarly for the  shifting $p\to p+K_1+K_2$.
Above relations \eref{Tri-gen-shift-1-1} by shifting can be summarized as the
cyclic symmetry $Z_3$ of three ordered lists
\bea g_3: \{(n_1,n_2, n_3);(K_1, K_2, K_3);(M_1, M_2,m_1)\} \to \{(n_2,n_3,n_1);(K_2, K_3, K_1); (M_2, m_1, M_1)\}~~~\label{Tri-gen-sym-1}  \eea
We can also consider the variable changing $p\to -p$ in \eref{Tri-gen-1-1} to get\footnote{
When changing $p\to \W p=-p$, the
$\int_{-\infty}^{+\infty} dp\to \int_{+\infty}^{-\infty} -d\W p= \int_{-\infty}^{+\infty} d\W p$, i.e., the measure is invariant under the reflection.  }
\bea I_3(n_1,n_2,n_3)[K_1,K_2,K_3;M_1, M_2, m_1]& = & \int\frac{d^{4-2\epsilon}p}{(2\pi)^{4-2\epsilon}}\frac{1}{(p^2-M_1^2)^{n_1}
((p+K_1)^2-M_2^2)^{n_2}((p-K_3)^2-m_1^2)^{n_3}}\nn
& = &I_3(n_1,n_3,n_2)[K_3,K_2,K_1;M_1, m_1, M_2] ~~~\label{Tri-gen-shift-2-1} \eea
which  can be summarized as the
reflection symmetry $Z_2$ of three ordered lists
\bea g_2: \{(n_1,n_2, n_3);(K_1, K_2, K_3);(M_1, M_2,m_1)\} \to \{(n_1,n_3, n_2);(K_3, K_2, K_1); (M_1, m_1, M_2)\} ~~~\label{Tri-gen-sym-2}  \eea
When combining $g_2,g_3$ together, we generate the permutation group $S_3$. Using the symmetric property, we can connect the reduction of one integral to another integral. For example, we know the expansion of $I_{3}(1,1,n)[K_1,K_2, K_3;M_1, M_2, m_1]$ and we want to calculate the expansion of
$I_{3}(1,n,1)[K_1, K_2,K_3;M_1, M_2, m_1]$. Using \eref{Tri-gen-shift-1-1}, we have
\bea & & I_{3}(1,n,1)[K_1, K_2,K_3;M_1, M_2, m_1]  =  g_3 \left\{ I_{3}(1,1,n)[K_3, K_1,K_2;m_1,M_1, M_2]\right\} \nn
& = & g_3 \left\{ c_{3\to 3}(n_1, n_2, n_3) {\cal I}_{3} +
\sum_{i=1}^3 c_{3\to 2;\bar{i}}(n_1, n_2, n_3) {\cal I}_{2;\bar{i}}+ \sum_{i=1}^3 c_{3 \to 1;i}( n_1, n_2, n_3) {\cal I}_{1;i}\right\}~.~~\label{Tri-gen-sym-3-1} \eea
Noticing that ${\cal I}_3[K_1, K_2,K_3;M_1, M_2, m_1]= g_3\left\{ {\cal I}_3[K_3, K_1,K_2;m_1,M_1, M_2]\right\}$, we have
\bea c_{3\to 3}(1,n,1)[K_1,K_2, K_3;M_1, M_2, m_1]  =  g_3 \left\{c_{3\to 3}(1,1,n)[K_3, K_1,K_2;m_1,M_1, M_2]\right\} ~.~~\label{Tri-gen-sym-3-2} \eea
Similarly, for bubble part we have ${\cal I}_{2;\bar{i}}[K_1,K_2, K_3;M_1, M_2, m_1]  =   g_3\left\{ {\cal I}_{2;\bar{i}}[K_3, K_1,K_2;m_1,M_1, M_2]\right\}$, thus
\bea  c_{3\to 2;\bar{i}}(1,n,1)[K_1,K_2, K_3;M_1, M_2, m_1]  =   g_3\left\{ c_{3\to 2;\bar{i}}(1,1,n)[K_3, K_1,K_2;m_1,M_1, M_2]\right\}~.~~\label{Tri-gen-sym-3-3} \eea
One simple consequence of above symmetry property is that if we know the expansion of $I_3(1,1,2)$, we know also $I_3(1,2,1)$ and $I_3(2,1,1)$.

Now we show how to use $I_3(1,1,2)$, $I_3(1,2,1)$ and $I_3(2,1,1)$ to get the expansion of general $I_3(n_1,n_2,n_3)$ using only differentiation.  Let us start from
$I_{3}(1,1,n_3)$ first. It is easy to see that
\bea I_{3}(1,1,n_3)
& = &{1\over (n_3-1)} {d\over d(m_1^2)} I_{3}(1,1,n_3-1) \nn
& = & {1\over (n_3-1)} {d\over d(m_1^2)} \left\{ c_{3\to 3}(1,1, n_3-1) {\cal I}_{3} +
\sum_{i=1}^3 c_{3\to 2;\bar{i}}(1,1, n_3-1) {\cal I}_{2;\bar{i}}+...\right\} \nn
& = & {1\over (n_3-1)} {dc_{3\to 3}(1,1, n_3-1)\over d(m_1^2)}  {\cal I}_{3}+ {c_{3\to 3}(1,1, n_3-1)\over (n_3-1)} I_3(1,1,2)+\sum_{i=1}^3 {dc_{3\to 2;\bar{i}}(1,1, n_3-1)\over (n_3-1) d(m_1^2)}{\cal I}_{2;\bar{i}} \nn
& & + {c_{3\to 2;\bar{1}}(1,1, n_3-1)\over (n_3-1)} I_{2;\bar{1}}(1,2)
+ {c_{3\to 2;\bar{2}}(1,1, n_3-1)\over (n_3-1)} I_{2;\bar{2}}(2,1)+...~~~\label{Tri-rec-1-1}\eea
where in the last line, since the ${\cal I}_{2;\bar{3}}$ does not depend on $m_1$, the action of ${d\over dm_1^2}$ is zero (see \eref{Tri-gen-3-2}, \eref{Tri-gen-3-3} and \eref{Tri-gen-3-4}).
Now, using the expansion of $I_3(1,1,2)$, $I_2(1,2)$ and  $I_2(2,1)$, we get immediately
the recurrence relation
\bea c_3(1,1,n_3) & = & {1\over (n_3-1)} {dc_{3\to 3}(1,1, n_3-1)\over d(m_1^2)}
+{c_{3\to 3}(1,1, n_3-1)\over (n_3-1)} c_{3\to 3}(1,1,2) \nn
c_{3\to 2;\bar{1}}(1,1,n_3) & = & {c_{3\to 3}(1,1, n_3-1)\over (n_3-1)}c_{3\to 2;\bar{1}}(1,1,2)+{1\over (n_3-1)} {dc_{3\to 2;\bar{1}}(1,1, n_3-1)\over d(m_1^2)}\nn & & +{c_{3\to 2;\bar{1}}(1,1, n_3-1)\over (n_3-1)} c_{2\to 2;\bar{1}}(1,2)\nn
c_{3\to 2;\bar{2}}(1,1,n_3) & = & {c_{3\to 3}(1,1, n_3-1)\over (n_3-1)}c_{3\to 2;\bar{2}}(1,1,2)+{1\over (n_3-1)} {dc_{3\to 2;\bar{2}}(1,1, n_3-1)\over d(m_1^2)}\nn & & +{c_{3\to 2;\bar{2}}(1,1, n_3-1)\over (n-1)} c_{2\to 2;\bar{2}}(2,1)\nn
c_{3\to 2;\bar{3}}(1,1,n_3) & = & {c_{3\to 3}(1,1, n_3-1)\over (n_3-1)}c_{3\to 2;\bar{3}}(1,1,2)+{1\over (n_3-1)} {dc_{3\to 2;\bar{3}}(1,1, n_3-1)\over d(m_1^2)}~~~\label{Tri-rec-1-2}\eea
Knowing \eref{Tri-rec-1-2}, it is easy to use \eref{Tri-gen-sym-1} and \eref{Tri-gen-sym-2}
to get reduction coefficients for $I_3(1,n,1)$ and $I_3(n,1,1)$ as shown in \eref{Tri-gen-sym-3-1}. One point we want to emphasize
is that since  $g_2, g_3$ acts on $K,M$ also, the kinematic dependence of $K,M$ in \eref{Tri-rec-1-2} should be carefully identified  although for simplicity we have not written  them down explicitly.

Knowing the reduction of $I_3(1,1,n_3)$, now we consider the case $I_3(1,n_2,n_3)$. Similarly to
\eref{Tri-rec-1-1}, we have
\bea
I_3(1,n_2,n_3)
&=&\frac{1}{(n_2-1)}\frac{\d}{\d M_2^2}\left\{c_{3\to 3}(1,n_2-1,n_3)I_3+\sum_{i=1}^3 c_{3\to 2;\bar i}(1,n_2-1,n_3)I_{2;\bar i}+\cdots\right\}
~~~\label{Tri-rec-2-1}\eea
% %
thus we can get the recurrence relation for reduction coefficients
% %
\bea
c_{3\to 3}(1,n_2,n_3)&=&\frac{1}{n_2-1}\left(\frac{\d c_{3\to 3}(1,n_2-1,n_3)}{\d M_2^2}+c_{3\to 3}(1,n_2-1,n_3)c_{3\to 3}(1,2,1)\right)\nn
c_{3\to 2;\bar 1}(1,n_2,n_3)&=&\frac{1}{n_2-1}\Big(c_{3\to 3}(1,n_2-1,n_3)c_{3\to 2;\bar 1}(1,2,1)+\frac{\d c_{3\to 2;\bar 1}(1,n_2-1,n_3)}{\d M_2^2}\nn
&&~~+c_{3\to 2;\bar 1}(1,n_2-1,n_3)c_{2\to 2,\bar{1}}(2,1)\Big)\nn
c_{3\to 2;\bar 2}(1,n_2,n_3)&=&\frac{1}{n_2-1}\Big(c_{3\to 3}(1,n_2-1,n_3)c_{3\to 2;\bar 2}(1,2,1)+\frac{\d c_{3\to 2;\bar 2}(1,n_2-1,n_3)}{\d M_2^2}\Big)\nn
c_{3\to 2;\bar 3}(1,n_2,n_3)&=&\frac{1}{n_2-1}\Big(c_{3\to 3}(1,n_2-1,n_3)c_{3\to 2;\bar 3}(1,2,1)+\frac{\d c_{3\to 2;\bar 3}(1,n_2-1,n_3)}{\d M_2^2}\nn
&&+c_{3\to 2;\bar 3}(1,n_2-1,n_3)c_{2\to 2}(1,2)\Big)~~~\label{Tri-rec-2-2}
\eea
% %
So if we have known the expansion coefficients of $I_3(1,n_2-1,n_3)$ to the basis $I_3$
and $I_2$, we can derive the expansion of  $I_3(1,n_2,n_3)$.

Finally, for $I_3(n_1,n_2,n_3)$ with $n_1>1$, similar action
lead to the recurrence relation
\bea
c_{3\to 3}(n_1,n_2,n_3)&=&\frac{1}{n_1-1}\Big(\frac{\d c_{3\to 3}(n_1-1,n_2,n_3)}{\d M_1^2}+c_{3\to 3}(n_1-1,n_2,n_3)c_{3\to 3}(2,1,1)\Big)\nn
c_{3\to 2;\bar 1}&=&\frac{1}{n_1-1}\Big(c_{3\to 3}(n_1-1,n_2,n_3)c_{3\to 2;\bar 1}(2,1,1)+\frac{\d c_{3\to 2;\bar 1}(n_1-1,n_2,n_3)}{\d M_1^2}\Big)\nn
c_{3\to 2;\bar 2}&=&\frac{1}{n_1-1}\Big(c_{3\to 3}(n_1-1,n_1,n_3)c_{3\to 2;\bar 2}(2,1,1)+\frac{\d c_{3\to 2;\bar 2}(n_1-1,n_2,n_3)}{\d M_1^2}\nn
&&+c_{3\to 2;\bar 2}(n_1-1,n_2,n_3)c_{2\to 2;\bar{2}}(2,1)\Big)\nn
c_{2;\bar 3}&=&\frac{1}{n_1-1}\Big(c_{3}(n_1-1,n_2,n_3)c_{2;\bar 3}(2,1,1)+\frac{\d c_{2;\bar 3}(n_1-1,n_2,n_3)}{\d M_1^2}\nn
&&+c_{2;\bar 3}(n_1-1,n_2,n_3)c_{2,1}[K_1,M_1,M_2]\Big)
\eea

%%%%%%%%%%%%%%%%%%%%%
\subsection{Triangle(1,1,2)}
%%%%%%%%%%%%%%%%%%%%%%%

From the recurrence relation in previous subsection, we see that all computations have been reduced to the reduction of $I_3(1,1,2)$. Although there are other methods to reduce $I_3(1,1,2)$ such as the IBP method. In this part, we show how to use unitarity cut method to fulfill the task. Using our idea, we can write
\bea I_3(1,1,2)={\d \over \d m_1^2} I_{3}(1,1,1).~~~~\label{Tri-112-1-1}\eea
To get all reduction coefficients, we need to calculate all three cuts. However, different cuts can be related using the symmetry discussed in previous subsection, thus we will focus on only one cut. 

%Now we use three different cuts to calculate the right hand side (RHS) to get all reduction coefficients.

%%%%%%%%%%%%%%%%%%%%%
\subsubsection{Cut $K_1$}
%%%%%%%%%%%%%%%%%%%%%

For this cut, the RHS is given by
\bea {\cal C}_{K_1}(I_3(1,1,2))= {\d \over \d m_1^2} {\cal C} (I_3)= -(\frac{4K_1^2}{\Delta[K_1,M_1,M_2]})^{\epsilon}\frac{1}{\sqrt{\Delta_{3;m=0}}}\frac{\partial}{\partial m_1^2}Tri^{(0)}(Z)~~~~\label{Tri-112-2-1}
\eea
where we have used the fact that only $Z$ contains the $m_1$. Carrying out the derivative, we get
\bea
\frac{\partial}{\partial m_1^2}Tri^{(0)}(Z)&= & \frac{\partial Z}{\partial m_1^2}\frac{\partial}{\partial Z}Tri^{(0)}(Z),~~~~~~
\frac{\partial Z}{\partial m_1^2}=  \frac{2K_1^2}{\sqrt{\Delta_{3;m=0}\Delta[K_1,M_1,M_2]}} \nn
 \frac{\d}{\d Z} Tri^{(0)}(Z)& =&\int_0^1duu^{-1-\epsilon}\Big[\frac{1}{Z+\sqrt{1-u}}-\frac{1}{Z-\sqrt{1-u}}\Big]\nn
& =&2\int_0^1duu^{-1-\epsilon}\frac{1-u}{\sqrt{1-u}(1-u-Z^2)}~~~~\label{Tri-112-2-3}\eea
Among two terms of $\frac{\d}{\d Z} Tri^{(0)}(Z)$, by comparing with \eref{Tri-gen-2-2}, the second term is
\bea 2\int_0^1duu^{-1-\epsilon}\frac{-u}{\sqrt{1-u}(1-u-Z^2)}
=-\frac{2\epsilon}{Z}Tri^{(0)}(Z)~~~~\label{Tri-112-2-4}\eea
For the first term, noting that
\bea & &
\int_0^1duu^{-1-\epsilon}\frac{1}{\sqrt{1-u}}=(1-Z^2)\int_0^1duu^{-1-\epsilon}\frac{1}{\sqrt{1-u}(1-u-Z^2)}-\int_0^1
duu^{-\epsilon}\frac{1}{\sqrt{1-u}(1-u-Z^2)}~~~~\label{Tri-112-2-5}
\eea
we have
\bea \frac{1}{1-Z^2}\frac{\epsilon}{Z}Tri^{(0)}(Z)+\frac{1-2\epsilon}{1-Z^2}Bub^{(0)} ~~~~\label{Tri-112-3-1}\eea
Putting all together, we get
\bea
\frac{\d}{\d Z}Tri^{(0)}(Z)&= & \frac{2(1-2\eps)}{1-Z^2}Bub^{(0)}+\frac{2Z\eps}{1-Z^2}Tri^{(0)}(Z)~~~~\label{Tri-112-3-2}
\eea
By identifying
\bea
\frac{\partial}{\partial m_1^2}CutI_3(1,1,1)=c_{3\to 3;K_1}(1,1,2) CutI_3(1,1,1)+c_{3\to 2;\bar{3};K_1}(1,1,2)Cut I_2(1,1)~~~~\label{Tri-112-3-4}
\eea
we get the reduction coefficients as
\bea
c_{3\to 3;K_1}(1,1,2) &= &\frac{4K_1^2}{\sqrt{\Delta_{3;m=0}\Delta[K_1,M_1,M_2]}}\frac{Z\epsilon}{1-Z^2}
~~~~\label{Tri-112-4-1-0}\\
c_{3\to 2;\bar{3};K_1}(1,1,2)&= & -\frac{4K_1^2}{\Delta[K_1,M_1,M_2]\Delta_{3;m=0}}\frac{1-2\epsilon}{1-Z^2} ~~~~~~~\label{Tri-112-4-1}\eea
where for simplicity we have not expanded $Z$ further (see  \eref{Tri-gen-2-3}).
We want to remark that the coefficients $c_{3\to 3;K_1}$ and $c_{3\to 2;\bar{3};K_1}$ are nothing, but the $c_{3\to 3}$ and $c_{3\to 2;\bar{3}}$ is the reduction
\bea
I_3(1,1,2)=c_{3\to 3}(1,1,2)I_3+c_{3\to 2;\bar{3}}(1,1,2)I_{3;\bar{3}}+\cdots\cdots ~~~~\label{Tri-112-4-2}
\eea
The reason that we have added subscript $K_1$ in \eref{Tri-112-4-1-0} and \eref{Tri-112-4-1} is
to emphasize that these two expressions are calculated using the cut $K_1$.

%%%%%%%%%%%%%%%%%
\subsubsection{Cut $K_2$}
%%%%%%%%%%%%%%%%%%%%

No we consider the cut $K_2$. The use the cut result \eref{Tri-gen-2-1}, we need to
rewrite the form \eref{Tri-gen-1-1} into the standard form for the cut $K_2$ by shifting the integral momentum $p\to p+K_1$, thus
it becomes
\bea & & I_3(n_1,n_2,n_3)[K_1,K_2,K_3;M_1, M_2, m_1]=\int\frac{d^{4-2\epsilon}p}{(2\pi)^{4-2\epsilon}}\frac{1}{((p+K_1)^2-M_1^2)^{n_1}
(p^2-M_2^2)^{n_2}((p-K_2)^2-m_1^2)^{n_3}}\nn
& = &I_3(n_2,n_3,n_1)[K_2,K_3,K_1;M_2,m_1,M_1]=g_3\{I_3(n_1,n_2,n_3)[K_1,K_2,K_3;M_1,M_2,m_1]\} ~~~\label{Tri-gen-K2-1-1-1} \eea
For the case $n_3=2$, $n_1=n_2=1$
\bea & & I_3(1,1,2)[K_1,K_2,K_3;M_1, M_2, m_1] = {\d \over \d m_1^2} I_3(1,1,1)[K_1,K_2,K_3;M_1, M_2, m_1]\nn
&= & {\d \over \d m_1^2}I_3(1,1,1)[K_2,K_3,K_1;M_2,m_1,M_1]= g_3\{\frac{\d}{\d M_2^2}I_3(1,1,1)[K_1,K_2,K_3;M_1,M_2,m_1]\}~~~~\label{K2-to-K1} \eea
Thus we have reduced the problem to the cut $K_1$, but with different action ${\d \over \d M_2^2}$.

For the action of $\frac{\d}{\d M_2^2}$, there are two terms containing the parameter $M_2$ in the $Cut_{K_1}(I_3(1,1,2))$.
\bea
\frac{\d}{\d M_2^2}Cut_{K_1}(I_{3}(1,1,2))=&-\frac{1}{\sqrt{\Delta_{3;m=0}}}\Big(\frac{\partial}{\partial M_2^2}(\frac{\Delta}{4K_1^2})^{-\epsilon}\Big)Tri^{(0)}(Z)+(-\frac{1}{\sqrt{\Delta_{3;m=0}}})(\frac{\Delta}{4K_1^2})^{-\epsilon}\frac{\partial}{\partial M_2^2}Tri^{(0)}(Z)~~~\label{Tri-112-K2-2-1}
\eea
where the $\Delta$ means $\Delta[K_1,M_1,M_2]$. The first term in \eref{Tri-112-K2-2-1} is trivial, while the derivative part in the second term is just $\frac{\d Z}{\d M_2^2}\frac{\d }{\d Z}Tri^{(0)}(Z)$,
where $\frac{\d }{\d Z}Tri^{(0)}(Z)$ has been given in \eref{Tri-112-3-2}.
Put them altogether, we get
\bea
\frac{\d}{\d M_2^2}C_{K_1}(I_{3})=\widetilde c_{3}C(I_3)+\widetilde c_{2}C(I_2)~~~\label{Tri-112-K2-3-1}
\eea
with
\bea
\tilde c_{3} &= &\Delta^{\eps}\frac{\d}{\d M_2^2}\Delta^{-\eps}+\frac{\d Z}{\d M_2^2}\frac{2Z\eps}{1-Z^2},~~~~
\tilde c_{2} =  -\frac{1}{\sqrt{\Delta_{3;m=0}}}\frac{K_1^2}{\sqrt{\Delta}}\frac{\d Z}{\d M_2^2}\frac{2(1-2\eps)}{1-Z^2}~~~\label{Tri-112-K2-3-2}
\eea
Of course, the $\tilde c_2$ and $\tilde c_{3}$ in \eref{Tri-112-K2-3-2} are not the final result, since we have to do the permutation $g_3$ to get
$c_{3\to 3}(1,1,2)=g_3(\W c_2)$ and $c_{3\to 2;\bar{1}}(1,1,2)=g_3(\W c_2)$.  One can find that the $g_3(\W c_2)$ is equal to  $c_{3\to 3}(1,1,2)$ given in \eref{Tri-112-4-1-0}, thus we have passed the first consistent check.

%%%%%%%%%%%%%%%%%
\subsubsection{Cut $K_3$}
%%%%%%%%%%%%%%%%%%%%

Now we consider the cut $K_3$. Again we need to rewrite the form \eref{Tri-gen-1-1} into the standard form for the cut $K_3$ by shifting the integral momentum $p\to p+K_1+K_2$.
With similar argument we will have
\bea & & I_3(1,1,2)[K_1,K_2,K_3;M_1, M_2, m_1] = \frac{\d}{\d m_1^2}I_3(1,1,1)[K_3,K_1,K_2;m_1,M_1,M_2]\nn &=&g_3^{-1}\{\frac{\d}{\d M_1^2}I_3(1,1,2)[K_1,K_2,K_3;M_1,M_2,m_1]\}~~~~\label{K3-to-K1} \eea
thus the cut $K_3$ has been reduced to the cut $K_1$ with the action ${\d\over \d M_1^2}$.
The calculation is similar to the previous subsection.
After simplification, we have
% %
\bea
\frac{\d}{\d M_1^2}C_{K_1}(I_3)&=&\WH c_3C_{K_1}(I_3)+\WH c_{2}C_{K_1}(I_2)
\eea
% %
with the coefficients
% %
\bea
\WH c_{3}&=&-\eps\Delta^{-1}\frac{\d\Delta}{\d M_1^2}+\frac{\d Z}{\d M_1^2}\frac{2Z\eps}{1-Z^2}\nn
\WH c_{2}&=&(-\frac{1}{\sqrt{\Delta_{3;m=0}}})\frac{K_1^2}{\sqrt{\Delta}}\frac{\d Z}{\d M_1^2}\frac{2(1-2\eps)}{1-Z^2}
\eea
% %
After doing the permutation $g_3^{-1}$ on the coefficients above, we get the $c_{3\to 3}$ and $c_{3\to 2;\bar 2}$ in this cut. Again, one can check the $c_{3\to 3}$ is the same with $c_{3\to 3;K_1}$ in \eref{Tri-112-4-1-0}.

For the case $I_3(1,1,2)$ there is another consistent check we can do. Noticing that when we do
following changing variable $p\rightarrow -p+K_1$,  the \eref{Tri-gen-1-1} becomes to
% %
\bea
I_3(n_1,n_2,n_3)[1]=\int\frac{d^{4-2\eps}p}{(2\pi)^{4-2\eps}}\frac{1}
{((p-K_1)^2-M_1^2)^{n_1}(p^2-M_2^2)^{n_2}((p+K_2)-m_1^2)^{n_3}}
\eea
% %
Thus for $(n_1,n_2,n_3)=(1,1,2)$, there is the symmetry among the reduction coefficients, i.e., under the the permutation $K_2\longleftrightarrow K_3,M_1\longleftrightarrow M_2$,
we should have $c_{3\to 3}$ invariant and $c_{3\to 2;\bar 2}\leftrightarrow c_{3\to 2;\bar 1}$. This can be easily checked by MATHEMATICA for our results.

%%%%%%%%%%%%%%%%%%%%
\subsubsection{A short summary}
%%%%%%%%%%%%%%%%%%%%%
In the subsection, we have reduced the $I_3$(1,1,2) to triangle $I_3$ and bubble $I_2$. After taking three different cuts, we get the all needed triangle and bubble coefficients of the reduction of $I_3(1,1,2)$
\bea
I_3(1,1,2)=c_{3\to 3}(1,1,2)I_3+\sum_{i=1}^3c_{3\to 2;\bar i}(1,1,2)I_{2;\bar i}+\cdots~~~~~~~~~~~\label{Tri-112-com}
\eea
where the $\cdots$ represents the tadpoles neglected in the whole paper.
The coefficients in \eref{Tri-112-com} are
%%%
\bea
c_{3\to 3}(1,1,2)&=&\frac{4K_1^2}{\sqrt{\Delta_{3;m=0}\Delta[K_1,M_1,M_2]}}\frac{Z\eps}{1-Z^2}\nn
c_{3\to 2;\bar 3}(1,1,2)&=&-\frac{4K_1^4}{\Delta[K_1,M_1,M_2]\Delta_{3;m=0}}\frac{1-2\eps}{1-Z^2}\nn
c_{3\to 2;\bar 1}(1,1,2)&=&\hat g_{3}\Big(-\frac{1}{\sqrt{\Delta_{3;m=0}}}\frac{K_1^2}{\sqrt{\Delta[K_1,M_1,M_2]}}\frac{\d Z}{\d M_2^2}\frac{2(1-2\eps)}{1-Z^2}\Big)\nn
c_{3\to 2;\bar 2}(1,1,2)&=&\hat g_{3}^{-1}\Big(-\frac{1}{\sqrt{\Delta_{3;m=0}}}\frac{K_1^2}{\sqrt{\Delta[K_1,M_1,M_2]}}\frac{\d Z}{\d M_1^2}\frac{2(1-2\eps)}{1-Z^2}\Big)
\eea
%%%
The result is confirmed with IBP method using LiteRed \cite{Lee:2013mka}.
We have also carried out the reduction of $I_3(1,2,1)$, $I_3(2,1,1)$ and $I_3(1,2,2)$ using the method laid out in the paper and found perfect match with the IBP method.

%%%%%%%%%%%%%%%%%%%%%%%%%%%%%%%%%%%% %%%%%%%
\section{Box}
%%%%%%%%%%%%%%%%%%%%%%%%%%%%%%%%%%

For the box topology, we define

%%%
\bea
I_4(n_1,n_2,n_3,n_4)=\int\frac{d^{4-2\eps}p}{(2\pi)^{4-2\eps}}\frac{1}{(p^2-M_1^2)^{n_1}
((p-K_1)^2-M_2^2)^{n_2}((p-K_1-K_2)^2-m_1^2)^{n_3}((p+K_4)^2-m_2^2)^{n_4}}~~~~\label{Box-gen-1-1}
\eea
%%%
or
%%%
\bea
I_4(n_1,n_2,n_3,n_4)=\int\frac{d^{4-2\eps}p}{(2\pi)^{4-2\eps}}\frac{1}
{(p^2-M_1^2)^{n_1}((p-K)^2-M_2^2)^{n_2}((p-P_1)^2-m_1^2)^{n_3}((p-P_2)^2-m_2^2)^{n_4}}
~~~~\label{box-basis}
\eea
%%%
for later convenience,
where we have labeled the cut momentum as $K$ and   the masses of two cut propagators  as $M_1$ and $M_2$ respectively. For the other two momenta we denote them as $P_1$ and $P_2$ respectively. For example, if we choose $K_1$ as the cut momentum, we will have  $P_1=K_1+K_2$ and $P_2=-K_4$, but  if the cut momentum is $K_1+K_2$, we will have $P_1=K_1, P_2=K_1+K_2+K_3$.

The special case with $n_i=1, i=1,2,3,4$ gives the scalar basis, which we will denote as $I_4$. The unitarity cut of $I_4$ with the cut momentum $K$ has been given by \cite{Britto:2006fc}
%%%%%
\bea
C(I_4)=(\frac{\Delta}{4K^2})^{-\eps}\frac{b}{2K^2}\int_0^1duu^{-1-\eps}\frac{1}
{\sqrt{B-Au}}ln(\frac{D-Cu+\sqrt{1-u}\sqrt{B-Au}}{D-Cu-\sqrt{1-u}\sqrt{B-Au}})~~~\label{Box-cut-1-1}
\eea
%%%%
with the parameters $\Delta$ in \eref{Bu-gen-1-3} and\footnote{The $A$ corresponds to the second-type singularity of box.}
%%%
\bea
A&=&-\frac{b^4}{K^2}det
\left(\begin{array}{ccc}
P_1^2 & P_1\cdot P_2 &P_1\cdot K\\
P_1\cdot P_2 &P_2^2&P_2\cdot K \\
P_1\cdot K&P_2\cdot K&K^2
\end{array}\right),~~~~~
%%%
B=-det
\left(\begin{array}{cc}
R_1^2&R_1\cdot R_2\\
R_1\cdot R_2&R_2^2\\
\end{array}\right)\nn
C&=&\frac{b^2}{K^2}det
\left(\begin{array}{cc}
P_1\cdot P_2&P_1\cdot K\\
P_2\cdot K&K^2\\
\end{array}
\right),~~~~~~~~~
D=R_1\cdot R_2~~~\label{Box-cut-1-2}
\eea
%%%
where
%%%
\bea
b&=&\frac{\sqrt{\Delta[K,M_1,M_2]}}{K^2},~~~~
a_i=\frac{P_i^2+M_1^2-m_1^2}{K^2},~~~
R_i=-bP_i+\Big(a_i-\frac{P_i\cdot K}{K^2}(a-b)\Big)K~~~\label{Box-cut-1-3}
\eea
%%%
With the definition
\bea
Box^{(n)}&=&\int_0^1duu^{-1-\eps}\frac{u^n}{\sqrt{B-Au}}
ln(\frac{D-Cu+\sqrt{1-u}\sqrt{B-Au}}{D-Cu-\sqrt{1-u}\sqrt{B-Au}})~~~\label{Box-cut-2-1}
\eea
we can write $C(I_4)$ as
\bea
C(I_4)=(\frac{\Delta}{4K^2})^{-\eps}\frac{b}{2K^2}Box^{(0)}~~\label{def}
\eea
and there is also a recurrence relation of $Box^{(n)}$, which is given by \cite{Anastasiou:2006gt,Britto:2006fc}
%%%
\bea
Box^{(n)}=\frac{n-1-\eps}{n-\frac{1}{2}-\eps}\frac{B}{A}Box^{(n-1)}-
\frac{n-1-\eps}{n-\frac{1}{2}-\eps}\frac{C_{Z_1}}{AZ_1}Tri^{(n-1)}(Z_1)-\frac{n-1-\eps}
{n-\frac{1}{2}-\eps}\frac{C_{Z_2}}{AZ_2}Tri^{(n-1)}(Z_2)~~~\label{Box-cut-2-2}
\eea
%%%
with the parameter
\bea
C_{Z_i}=D+(Z_{i}^2-1)C~~~\label{Box-cut-2-3}
\eea
To be calar, we list six possible cuts of a box, with $K_1$, $K_2$, $K_3$, $K_4$ in clockwise ordering. And there will be two cut triangles related to each  given cut momentum $K$ of box:
\begin{table}[htbp]
\centering
\label{table1}
\begin{tabular}{|c|c|c|c|c|}
\hline
Box Cut $K$ & $P_1$ & $P_2$ &Triangle One's ($K_1$,$K_3$)&Triangle Two's ($K_1$,$K_3$)\\
\hline
$K_1$ & $K_{12}$ & $-K_4$ & $(K_1,K_{34})$ & $(K_1,K_4)$ \\
\hline
$K_2$ & $K_{23}$ & $-K_1$ & $(K_2,K_{41})$ & $(K_2,K_1)$ \\
\hline
$K_3$ & $K_{34}$ & $-K_2$ & $(K_3,K_{12})$ & $(K_3,K_2)$\\
\hline
$K_4$ & $K_{41}$ & $-K_3$ & $(K_4,K_{23})$ & $(K_4,K_3)$\\
\hline
$K_{12}$ & $K_1$ & $-K_4$ & $(K_{34},K_2)$ & $(K_{12},K_4)$\\
\hline
 $K_{23}$ & $K_2$ & $-K_1$ & $(K_{41},K_3)$ & $(K_{23},K_1)$\\
\hline
\end{tabular}~~~~~~\label{Box-cut-2-4}\end{table}

~\\{\bf Symmetry analysis:}
Similar to the case of triangle, there are some symmetries of box  by momentum shifting and reflection. Let us define two generators
\bea
\hat g_4:&&\{((n_1,n_2,n_3,n_4);(K_1,K_2,K_3,K_4);(M_1,M_2,m_1,m_2)\}\nn
&&\to\{(n_2,n_3,n_4,n_1);(K_2,K_3,K_4,K_1);(M_2,m_1,m_2,M_1)\}~~~\label{Box-cut-3-4-1}
\eea
and
\bea
\hat g_2:&&\{((n_1,n_2,n_3,n_4);(K_1,K_2,K_3,K_4);(M_1,M_2,m_1,m_2)\}\nn
&&\to\{(n_1,n_4,n_3,n_2);(K_4,K_3,K_2,K_1);(M_1,m_2,m_1,M_2)\}~~~\label{Box-cut-3-4-2}
\eea
Using them we get the dihedral group $D_4$ and following relations
\bea
& & I_4(n_2,n_3,n_4,n_1)[K_2,K_3,K_4,K_1;M_2,m_1,m_2,M_1]=\WH g_4I_4(n_1,n_2,n_3,n_4)[K_1,K_2,K_3,K_4;M_1,M_2,m_1,m_2]\nn
& & I_4(n_3,n_4,n_1,n_2)[K_3,K_4,K_1,K_2;m_1,m_2,M_1,M_2]
=\WH g_4^2I_4(n_1,n_2,n_3,n_4)[K_1,K_2,K_3,K_4;M_1,M_2,m_1,m_2]\nn
& & I_4(n_4,n_1,n_2,n_3)[K_4,K_1,K_2,K_4;m_2,M_1,M_2,m_1]=\WH g_4^3I_4(n_1,n_2,n_3,n_4)[K_1,K_2,K_3,K_4;M_1,M_2,m_1,m_2]~~~~~~~\label{Box-cut-3-5}
\eea
and
\bea
&&I_4(n_1,n_4,n_3,n_2)[K_4,K_3,K_2,K_1;M_1,m_2,m_1,M_2]=\WH g_2I_4(n_1,n_2,n_3,n_4)[K_1,K_2,K_3,K_4;M_1,M_2,m_1,m_2] \nn
& & I_4(n_4,n_3,n_2,n_1)[K_3,K_2,K_1,K_4;m_2,m_1,M_2,M_1]=\WH g_4 \WH g_2I_4(n_1,n_2,n_3,n_4)[K_1,K_2,K_3,K_4;M_1,M_2,m_1,m_2] \nn
& & I_4(n_3,n_2,n_1,n_4)[K_2,K_1,K_4,K_3;m_1,M_2,M_1,m_2]=\WH g_4^2 \WH g_2I_4(n_1,n_2,n_3,n_4)[K_1,K_2,K_3,K_4;M_1,M_2,m_1,m_2] \nn
& & I_4(n_2,n_1,n_4,n_3)[K_1,K_4,K_3,K_2;M_2,M_1,m_2,m_1]=\WH g_4^3 \WH g_2I_4(n_1,n_2,n_3,n_4)[K_1,K_2,K_3,K_4;M_1,M_2,m_1,m_2]~~~~~~~~\label{Box-cut-3-6}
\eea
Furthermore, using the same idea in the subsection \ref{Tri-recur} we can write down
similar recurrence relation of reduction coefficients for general $I_4(n_1,n_2,n_3,n_4)$ using the expansion
of $I_4(2,1,1,1)$, $I_4(1,2,1,1)$, $I_4(1,1,2,1)$ and $I_4(1,1,1,2)$. However,
by relation \eref{Box-cut-3-5} and \eref{Box-cut-3-6}, all three cases $I_4(2,1,1,1)$, $I_4(1,2,1,1)$ and  $I_4(1,1,2,1)$ can be reduced to the reduction of   $I_4(1,1,1,2)$, thus
we need only to deal with the reduction of $I_4(1,1,1,2)$.

%%%%%%%%%%%%%%%%%%%%%%%%%
\subsection{Box $I_4(1,1,1,2)$ }
%%%%%%%%%%%%%%%%%%%%%%%%
Now, we use our method to calculate the reduction coefficients of $I_4(1,1,1,2)$. According our idea, we can write
\bea
I_4(1,1,1,2)=\frac{\d}{\d m_2^2}I_4(1,1,1,1)~~~~\label{box1112-1}
\eea
Comparing with the calculation of triangle, we need to consider  six different cuts:  four cuts with nearby propagators and two with
opposite propagators. By the symmetry, we could just calculate one for each type of cuts, and  get the others by proper permutation. To demonstrate our method, let us show the calculation for the cut $K_1$ only.  The reduction coefficients calculated for $I_4(1,1,1,2)$  has been checked using the LiteRed \cite{Lee:2013mka}.

For the cut $K_1$,  we have $K=K_1$, $P_1=K_1+K_2$ and $P_2=-K_4$ in \eref{box-basis}. To calculate $\frac{\d}{\d m_2^2}C(I_4)$, since the parameter $\Delta$ and $b$ do not contain the $m_2$, we could just write it as
\bea
\frac{\d}{\d m_2^2}C(I_4)=(\frac{\Delta}{4K^2})^{-\eps}\frac{b}{2K^2}\frac{\d}{\d m_2^2}Box^{(0)}
~~~~\label{box1112-K1-1}
\eea
 To simplify $\frac{\d}{\d m_2^2}Box^{(0)}$, first we  rewrite $Box^{(0)}$ as by partial integration and algebraic separation as
\bea
& & Box^{(0)}=\frac{2}{-A}\Big[(1+\eps)\int_0^1duu^{-2-\eps}\sqrt{B-Au}ln(\frac{D-Cu+\sqrt{1-u}\sqrt{B-Au}}{D-Cu-\sqrt{1-u}\sqrt{B-Au}})\nn
& &+\frac{\eps C_{Z_1}}{Z_1(Z_1^2-1)}Tri^{(0)}(Z_1)+\frac{\eps C_{Z_2}}{Z_2(Z_2^2-1)}Tri^{(0)}(Z_2)+\Big(\frac{C_{Z_1}(1-2\eps)}{Z_1^2-1}+\frac{C_{Z_2}(1-2\eps)}
{Z_2^2-1}\Big)Bub^{(0)}\Big]~~~\label{rebox2}
\eea
%%%
where the bubble and two triangles are specified by our cut $K$.
Now we consider the action of  $\frac{\d}{\d m_2^2}$. Since only $B$ and $D$ contains the parameter $m_2$, we can write directly
\bea
\frac{\d}{\d m_2^2}Box^{(0)}&=&\frac{2}{-A}\Big[(1+\eps)\int_0^1duu^{-2-\eps}\frac{B'_{m_2^2}}{2\sqrt{B-Au}}ln(\frac{D-Cu+\sqrt{1-u}\sqrt{B-Au}}{D-Cu-\sqrt{1-u}\sqrt{B-Au}}\nn
&&+(1+\eps)\int_0^1duu^{-2-\eps}\frac{(1-u)(u(2AD'_{(m_2^2)}-B'_{(m_2^2)}C)+(B'_{(m_2^2)}D-2BD'_{(m_2^2)}))}{\sqrt{1-u}((D-Cu)^2-(1-u)(B-Au))}\nn
& &+\frac{\d}{\d m_2^2}\Big(\frac{\eps C_{Z_1}}{Z_1(Z_1^2-1)}\Big)Tri^{(0)}(Z_1)+\frac{\d}{\d m_2^2}\Big(\frac{\eps C_{Z_2}}{Z_2(Z_2^2-1)}\Big)Tri^{(0)}(Z_2)\nn
& &+\frac{\eps C_{Z_1}}{Z_1(Z_1^2-1)}\frac{\d}{\d m_2^2}Tri^{(0)}(Z_1)+\frac{\eps C_{Z_2}}{Z_2(Z_2^2-1)}\frac{\d}{\d m_2^2}Tri^{(0)}(Z_2)\nn
& &+\frac{\d}{\d m_2^2}\Big(\frac{C_{Z_1}(1-2\eps)}{Z_1^2-1}+\frac{C_{Z_2}(1-2\eps)}{Z_2^2-1}\Big)Bub^{(0)}\Big]~~\label{box-2}
\eea
where the $B'_{(m_2^2)}$ and $D'_{(m_2^2)}$ means $\frac{\d}{\d m_2^2}B$ and $\frac{\d}{\d m_2^2}D$. There are five terms in the bracket. Since the third line and fifth line are just some simple differentiation over coefficients, we calculate the first, second and four lines only:
\begin{itemize}

\item (1) For the first term,  it is $\frac{(1+\eps)B'_{(m_2^2)}}{2}Box^{(-1)}$, while using using \eref{Box-cut-2-2},  we can expand $Box^{(-1)}$ as
\bea
Box^{(-1)}&=&\frac{\frac{1}{2}+\eps}{1+\eps}\frac{A}{B}Box^{(0)}+\frac{\eps}{1+\eps}\frac{C_{Z_1}}{BZ_1(1-Z_1^2)}Tri^{(0)}(Z_1)+\frac{\eps}{1+\eps}\frac{C_{Z_2}}{BZ_2(1-Z_2^2)}Tri^{(0)}(Z_2)\nn
& &+\frac{1-2\eps}{1+\eps}\Big(\frac{C_{Z_1}}{(1-Z_1^2)B}+\frac{C_{Z_2}}{(1-Z_2^2)B}\Big)Bub^{(0)}~~~\label{rebox}
\eea

\item (2) For the second term, using the trick of splitting terms and the recurrence relation,  we get
\bea
C_{3;Z_1}Tri^{(0)}(Z_1)+C_{3;Z_2}Tri^{(0)}(Z_2)+C_{2}Bub^{(0)}
\eea
with
\bea
C_{3;Z_1;K_1}&=&\frac{\eps Z_1\Big(\frac{\d B}{\d m_2^2}D-2B\frac{\d D}{\d m_2^2}+(Z_1^2-1)(\frac{\d B}{\d m_2^2}C-2A\frac{\d D}{\d m_2^2})\Big)}{\beta_1\beta_2(Z_1^2-1)^2(Z_1^2-Z_2^2)}\nn
C_{3;Z_2;K_2}&=&\frac{\eps Z_2\Big((\frac{\d B}{\d m_2^2}D-2B\frac{\d D}{\d m_2^2})+(Z_2^2-1)(\frac{\d B}{\d m_2^2}C-2A\frac{\d D}{\d m_2^2})\Big)}{\beta_1\beta_2(Z_2^2-1)^2(Z_2^2-Z_1^2)}\nn
C_{2;K_1}&=&\frac{n_{2;K_1}}{2\beta_1\beta_2(Z_1^2-1)^2(Z_2^2-1)^2(\eps+1)}~~\label{coe-123}
\eea
where
\bea
& & n_{2;K_1}=(2\eps-1)\Big(2\frac{\d B}{\d m_2^2}C(Z_1^2-1)(Z_2^2-1)(1+\eps)-4A\frac{\d D}{\d m_2^2}(Z_1^2-1)(Z_2^2-1)(1+\eps)\nn
&&+\frac{\d B}{\d m_2^2}D(-3-4\eps+Z_2^2(1+2\eps)+Z_1^2(1+Z_2^2+2\eps))-2B\frac{\d D}{\d m_2^2}(-3-4\eps+Z_2^2(1+2\eps)+Z_1^2(1+Z_2^2+2\eps)\Big)\nonumber
\eea

\item (3) For the fourth line with $\frac{\d}{\d m_2^2}Tri^{(0)}$, we could use result \eref{Tri-112-3-2} and \eref{Tri-112-2-3} to rewrite it as
\bean
-\frac{\d Z_1}{\d m_2^2}\frac{2\eps^2C_{Z_1}}{(Z_1^2-1)^2}Tri^{(0)}(Z_1)-\frac{\d Z_2}{\d m_2^2}\frac{2\eps^2C_{Z_2}}{(1-Z_2^2)^2}Tri^{(0)}(Z_2)
-\Big(\frac{\d Z_1}{\d m_2^2}\frac{2\eps(1-2\eps)C_{Z_1}}{Z_1(Z_1^2-1)^2}+\frac{\d Z_2}{\d m_2^2}\frac{2\eps(1-2\eps)C_{Z_2}}{Z_2(Z_2^2-1)^2}\Big)Bub^{(0)}
\eean

\end{itemize}

Now collecting all results together for \eref{box-2} we get our final result in the cut $K_1$:
\bea
\frac{\d}{\d m_2^2}I_4&=&c_{4\to 4;K_1}I_4+c_{4\to 3;\bar 4;K_1}I_{3;\bar 4}+c_{4\to 3,\bar 3;K_1}I_{3;\bar 3}I+c_{4\to 2;12;K_1}I_{2;12;K_1}+\cdots~~~~\label{box1112-K1-fin-1}
\eea
where
\bea
I_{3;\bar 4}&=&I_4(1,1,1,0),~~~
I_{3;\bar 3}=I_4(1,1,0,1),~~~
I_{2;12;K_1}=I_4(1,1,0,0)~~~~\label{box1112-K1-fin-1-1}
\eea
with the coefficients
\bea
c_{4\to 4;K_1}&=&-\frac{\frac{\d B}{\d m_2^2}(\frac{1}{2}+\eps)}{B}\nn
c_{4\to 3;\bar 4;K_1}&=&\frac{-b\sqrt{\Delta_{3;m=0}[Z_1]}}{2K_1^2}\Big(\frac{-\eps C_{Z_1}\frac{\d B}{\d m_2^2}}{ABZ_1(1-Z_1^2)}-\frac{2(1+\eps)}{A}C_{3;Z_1}-\frac{2}{A}\frac{\d}{\d m_2^2}\Big(\frac{\eps C_{Z_1}}{Z_1(Z_1^2-1)}\Big)+\frac{2}{A}\frac{\d Z_1}{\d m_2^2}\frac{2\eps^2C_{Z_1}}{(Z_1^2-1)^2}\Big)\nn
c_{4\to 3;\bar 3;K_1}&=&\frac{-b\sqrt{\Delta_{3;m=0}[Z_2]}}{2K_1^2}\Big(\frac{-\eps C_{Z_2}\frac{\d B}{\d m_2^2}}{ABZ_2(1-Z_2^2)}-\frac{2(1+\eps)}{A}C_{3;Z_2}-\frac{2}{A}\frac{\d}{\d m_2^2}\Big(\frac{\eps C_{Z_2}}{Z_2(Z_2^2-1)}\Big)+\frac{2}{A}\frac{\d Z_2}{\d m_2^2}\frac{2\eps^2C_{Z_2}}{(1-Z_2^2)^2}\Big)\nn
c_{4\to 2;12;K_1}&=&\frac{1}{2K_1^2}\Big[-(1-2\eps)\Big(\frac{\frac{\d B}{\d m_2^2}C_{Z_1}}{AB(1-Z_1^2)}+\frac{\frac{\d B}{\d m_2^2}C_{Z_2}}{AB(1-Z_2^2)}\Big)-\frac{2(1+\eps)}{A}C_{2}\nn
& &+\frac{2}{A}\Big(\frac{\d Z}{\d m_2^2}\frac{2\eps(1-2\eps)C_{Z_1}}{Z_1(Z_1^2-1)^2}+\frac{\d Z}{\d m_2^2}\frac{2\eps(1-2\eps)C_{Z_2}}{Z_2(Z_2^2-1)^2}\Big)-\frac{2}{A}\frac{\d}{\d m_2^2}\Big(\frac{C_{Z_1}(1-2\eps)}{Z_1^2-1}+\frac{C_{Z_2}(1-2\eps)}{Z_2^2-1}\Big)\Big]~~\label{cutk1}
\eea
where $C_{3;Z_1}$, $C_{3;Z_2}$ and $C_2$ given in Eq.\eref{coe-123}, and
\bea
\Delta_{3;m=0}[Z_1]&=&4((K_1\cdot K_{34})^2-K_1^2K_{34}^2),~~~~
\Delta_{3;m=0}[Z_2]=4((K_1\cdot K_4)^2-K_1^2K_4^2)
\eea

Having given details for the cut $K_1$, the computation of other cuts will be similar, as shown in the section of triangle.   For simplicity we will not list them one by one. All coefficients have been checked using LiteRed \cite{Lee:2013mka}.

%%%%%%%%%%%%%%%%%%%%%%%%%%%%%%%%%%%%%%%
\section{Pentagon}
%%%%%%%%%%%%%%%%%%%%%%%%%%%%%%%%%%%%%%

For the pentagon, let us define
\bea
& & I_5(n_1,n_2,n_3,n_4,n_5)[K_1,K_2,K_3,K_4, K_5;M_1,M_2, m_1, m_2, m_3]\nn
& = & \int\frac{d^{4-2\eps}p}{(2\pi)^{4-2\eps}}\frac{1}{(p^2-M_1^2)^{n_1}
((p-K_1)^2-M_2^2)^{n_2}((p-K_{12})^2-m_1^2)^{n_3}((p-K_{123})^2-m_2^2)^{n_4}
((p+K_5)^2-m_3^2)^{n_5}}\nn
& = & \int\frac{d^{4-2\eps}p}{(2\pi)^{4-2\eps}}\frac{1}{(p^2-M_1^2)^{n_1}((p-K)^2-M_2^2)^{n_2}
((p-P_1)^2-m_1^2)^{n_3}((p-P_2)^2-m_2^2)^{n_4}((p-P_3)^2-m_3^2)^{n_5}}
~~~~\label{Pen-gen-1-1}
\eea
%%%
where the second form is suitable for the discussion of unitarity cut with cut momentum $K$ in various cuts. The master basis of pentagon is given by $I_5\equiv I_5(1,1,1,1,1)$ and the cut part with $K$ is given by \cite{Britto:2006fc}
\bea
C(I_5)&=&(\frac{\Delta}{4K^2})^{-\eps}(-b)\int_0^1duu^{-1-\eps}\frac{\sqrt{1-u}}{(K^2)^2}
\Big(\frac{S[Q_3,Q_2,Q_1,K]}{4\sqrt{(Q_3\cdot Q_2)^2-Q_3^2Q_2^2}}\ln(\frac{Q_3\cdot Q_2-\sqrt{(Q_3\cdot Q_2)^2-Q_3^2Q_2^2}}{Q_3\cdot Q_2+\sqrt{(Q_3\cdot Q_2)^2-Q_3^2Q_2^2}})\nn
&+&\frac{S[Q_3,Q_1,Q_2,K]}{4\sqrt{(Q_3\cdot Q_1)^2-Q_3^2Q_1^2}}\ln(\frac{Q_3\cdot Q_1-\sqrt{(Q_3\cdot Q_1)^2-Q_3^2Q_1^2}}{Q_3\cdot Q_1+\sqrt{(Q_3\cdot Q_1)^2-Q_3^2Q_1^2}})\nn
&+&\frac{S[Q_2,Q_1,Q_3,K]}{4\sqrt{(Q_2\cdot Q_1)^2-Q_2^2Q_1^2}}\ln(\frac{Q_2\cdot Q_1-\sqrt{(Q_2\cdot Q_1)^2-Q_2^2Q_1^2}}{Q_2\cdot Q_1+\sqrt{(Q_2\cdot Q_1)^2-Q_2^2Q_1^2}})\Big)
\eea
where $
 Q_i= -\left( b \sqrt{1-u}
\right) P_i+{ P_i^2+M_1^2-m_i^2-2z(K\cdot P_i) \over K^2} K
$ (see also \eref{Box-cut-1-3})
 and $S[Q_3,Q_2,Q_1,K]$ is a rational function defined as follows
%
%%%
\bea
S[Q_j,Q_i,Q_k,K]={T_{jik}\over T_2},~~~T_{jik}&=&-8det
\left(\begin{array}{ccc}
Q_k\cdot K & Q_j\cdot K & Q_i\cdot K\nn
Q_j\cdot Q_k & Q_j^2 & Q_j\cdot Q_i\nn
Q_i\cdot Q_k & Q_j\cdot Q_i & Q_i^2
\end{array}\right),~~~
T_2=-4det
\left(\begin{array}{ccc}
Q_3^2&Q_2\cdot Q_3&Q_1\cdot Q_3\nn
Q_2\cdot Q_3&Q_2^2&Q_2\cdot Q_1\nn
Q_1\cdot Q_3&Q_2\cdot Q_1&Q_1^2\nn
\end{array}\right)~~~\label{Pen-def-1-1}
\eea
The form of $C(I_5)$ is like an addition of three dirrefent $C(I_4)$, with the common factor $\sqrt{1-u}$. To simplify further, noticing that
there is  a common factor $(1-u)^2$ between $T_{jik}$ and $T_2$,  we can define (please notice that
$T_{jik}=T_{ijk}$)
%%%
%
\bea
T_{2,re}&=&\frac{T_2}{(1-u)^2}= H_0+H_1u,~~~
T_{12}=\frac{T_{213}}{(1-u)^2},~~~
T_{13}=\frac{T_{312}}{(1-u)^2},~~~
T_{23}=\frac{T_{321}}{(1-u)^2}~~~\label{Pen-def-1-2}
\eea
where $T_{ij}$ is independent  of $u$ and  $T_{2,re}$ is a liner function of $u$.
Furthermore, with parameters
\bea
a=\frac{K^2+M_1^2-M_2^2}{K^2},~~~
b=\frac{\sqrt{\Delta[K,M_1,M_2]}}{K^2},~~~
a_{i}\equiv \frac{P_{i}^2+M_1^2-m_i^2}{K^2}~~~\label{Pen-def-1-3}
\eea
 we  define
\bea
\alpha_i\equiv a_{i}K^2-aP_{i}\cdot K,~~~
\beta_{i}\equiv b^2\Big(P_i^2-\frac{(P_i\cdot K)^2}{K^2}\Big),~~~
\gamma_{ij}\equiv b^2\Big((P_i\cdot P_j)^2-\frac{(P_i\cdot K)(P_j\cdot K)}{K^2}\Big)~~~\label{Pen-def-1-4}
\eea
thus $A,B,C,D$ defined in \eref{Box-cut-1-2} for box cut and \eref{Pen-def-1-2}
can be simplified as
\bea
A_{12}=\gamma_{12}^2-\beta_1\beta_2,~~~
B_{12}=\frac{-\alpha_2^2\beta_1-\alpha_1^2\beta_2+2\alpha_1\alpha_2
\gamma_{12}}{K^2}+\gamma_{12}^2-\beta_1\beta_2,~~~
C_{12}=\gamma_{12},~~
D_{12}=\gamma_{12}+\frac{\alpha_1\alpha_2}{K^2}~~~~~~\label{Pen-def-1-5}
\eea
and
\bea
H_1&=&4\beta_1\beta_2\beta_3+8\gamma_{12}\gamma_{13}\gamma_{23}-4\beta_3\gamma_{12}^2-4\beta_2\gamma_{13}^2-4\beta_1\gamma_{23}^2\nn
H_0&=&-H_1+\frac{4}{K^2}(\alpha_1^2A_{23}+\alpha_2^2A_{13}+\alpha_3^2A_{12})\nn
&&+8\Big((D_{23}-C_{23})(\beta_1C_{23}-C_{12}C_{13})+(D_{13}-C_{13})(\beta_2C_{13}-C_{12}C_{23})+(D_{12}-C_{12})(\beta_3C_{12}-C_{13}C_{23})\Big)\nn
T_{12}&=&8(\alpha_3A_{12}+\alpha_1\beta_2\gamma_{13}-\alpha_2\gamma_{12}\gamma_{13}+\alpha_2\beta_1\gamma_{23}-\alpha_1\gamma_{12}\gamma_{23})\nn
T_{23}&=&8(\alpha_1A_{23}+\alpha_2\beta_3\gamma_{12}+\alpha_3\beta_2\gamma_{13}-\alpha_3\gamma_{12}\gamma_{23}-\alpha_2\gamma_{13}\gamma_{23})\nn
T_{13}&=&8(\alpha_2A_{13}+\alpha_1\beta_3\gamma_{12}+\alpha_3\beta_1\gamma_{23}
-\alpha_1\gamma_{13}\gamma_{23}-\alpha_3\gamma_{12}\gamma_{13})~~~\label{para}
\eea
With above new notations, we can rewrite the expression of $C(I_5)$ as
\bea
& & C(I_5)=(\frac{\Delta}{4K^2})^{-\eps}\frac{b}{4(K^2)^2}\int_0^1duu^{-1-\eps}
\times\frac{1}{H_0+H_1u}
\Big\{\frac{T_{23}}{\sqrt{B_{23}-A_{23}u}}\ln(\frac{D_{23}-C_{23}u+\sqrt{1-u}
\sqrt{B_{23}-A_{23}u}}{D_{23}-C_{23}u-\sqrt{1-u}\sqrt{B_{23}-A_{23}u}})~~~\label{I5-cut}\\
&&+\frac{T_{13}}{\sqrt{B_{13}-A_{13}u}}\ln(\frac{D_{13}-C_{13}u+\sqrt{1-u}\sqrt{B_{13}
-A_{13}u}}{D_{13}-C_{13}u-\sqrt{1-u}\sqrt{B_{13}-A_{13}u}})
+\frac{T_{12}}{\sqrt{B_{12}-A_{12}u}}\ln(\frac{D_{12}-C_{12}u+\sqrt{1-u}\sqrt{B_{12}
-A_{12}u}}{D_{12}-C_{12}u-\sqrt{1-u}\sqrt{B_{12}-A_{12}u}})\Big\}\nonumber
\eea
We can see that in  \eref{I5-cut} only parameters $\a_i$ contains $m_i^2$, so
\bea
\frac{\d}{\d m_i^2}&=&\frac{\d}{\d \alpha_i}~\frac{\d \alpha_i}{\d a_i}~\frac{\d a_i}{\d m_i^2}=-\frac{\d}{\d \alpha_i}
\eea
For later convenience, we  define three functions
\bea
pen_{ij}^{(n)}&=&\int_0^1duu^{n-1-\eps}\frac{1}{H_0+H_1 u}\frac{1}{\sqrt{B_{ij}-A_{ij}u}}\ln(\frac{D_{ij}-C_{ij}u+\sqrt{1-u}\sqrt{B_{ij}-A_{ij}u}}{D_{ij}-C_{ij}u-\sqrt{1-u}\sqrt{B_{ij}-A_{ij}u}})\nn
&=&\int_0^1duu^{n-1-\eps}\frac{1}{H_0+H_{1}u}\frac{1}{\sqrt{B_{ij}-A_{ij}u}}\ln(*_{ij}),~~~*_{ij}\equiv\frac{D_{ij}-C_{ij}u+\sqrt{1-u}
\sqrt{B_{ij}-A_{ij}u}}{D_{ij}-C_{ij}u-\sqrt{1-u}\sqrt{B_{ij}-A_{ij}u}}~~~~~~~~~\label{Pen-def-1-6}
\eea
thus the $C(I_5)$ could be written as
\bea
C(I_5)&=&(\frac{\Delta}{4K^2})^{-\eps}\frac{b}{4(K^2)^2}
\Big(T_{23}pen_{23}^{(0)}+T_{13}pen_{13}^{(0)}+T_{12}pen_{12}^{(0)}\Big)~~\label{ci5}
\eea
From the definition of \eref{Pen-def-1-6}, one can easily establish the recursion relation
by rewriting $u= { (H_0+ H_1 u)- H_0\over H_1}$. The cancelation of denominator $H_0+H_1 u$ in
\eref{Pen-def-1-6} is nothing, but the cut of corresponding box (see \eref{Box-cut-2-1}).

Similar to other sections, with momentum shifting and reflection, we can define the action
\bea \WH g_5:  & &
\left\{(n_1,n_2,n_3,n_4,n_5);(K_1,K_2,K_3,K_4,K_5);(M_1,M_2,m_1,m_2,m_3)\right\}\nn
&\to & \left\{(n_2,n_3,n_4,n_5,n_1);(K_2,K_3,K_4,K_5,K_1);(M_2,m_1,m_2,m_3,M_1)\right\}\nn
\WH g_2:  & &
\left\{(n_1,n_2,n_3,n_4,n_5);(K_1,K_2,K_3,K_4,K_5);(M_1,M_2,m_1,m_2,m_3)\right\}\nn
&\to & \left\{(n_2,n_1,n_5,n_4,n_3);(K_1,K_5,K_4,K_3,K_2);(M_2,M_1,m_3,m_2,m_1)\right\}\eea
thus if we write ${\WH I}_{5}\equiv I_5(n_1,n_2,n_3,n_4,n_5)[K_1,K_2,K_3K_4,K_5;M_1,M_2,m_1,m_2,m_3]$ we will have
\bea I_5(n_2,n_3,n_4,n_5,n_1)[K_2,K_3,K_4,K_5,K_1;M_2,m_1,m_2,m_3,M_1] & = &  \WH g_5 {\WH I}_{5} \nn
I_5(n_3,n_4,n_5,n_1,n_2)[K_3,K_4,K_5,K_1,K_2;m_1,m_2,m_3,M_1,M_2]& = &  \WH g_5^2 {\WH I}_{5} \nn
I_5(n_4,n_5,n_1,n_2,n_3)[K_4,K_5,K_1,K_2,K_3;m_2,m_3,M_1,M_2,m_1]& = &  \WH g_5^3 {\WH I}_{5}\nn
I_5(n_5,n_1,n_2,n_3,n_4)[K_5,K_1,K_2,K_3,K_4;m_3,M_1,M_2,m_1,m_2]& = &  \WH g_5^4 {\WH I}_{5}
~~~\label{Pen-sym-1}\eea
and
\bea I_5(n_2,n_1,n_5,n_4,n_3)[K_1,K_5,K_4,K_3,K_2;M_2,M_1,m_3,m_2,m_1]& = &  \WH g_2 {\WH I}_{5} \nn
I_5(n_3,n_2,n_1,n_5,n_4)[K_2,K_1,K_5,K_4,K_3;m_1,M_2,M_1,m_3,m_2]& = & \WH g_5^{-1} \WH g_2 {\WH I}_{5}\nn
I_5(n_4,n_3,n_2,n_1,n_5)[K_3,K_2,K_1,K_5,K_4;m_2,m_1,M_2,M_1,m_3]& = & \WH g_5^{-2} \WH g_2 {\WH I}_{5}\nn
I_5(n_5,n_4,n_3,n_2,n_1)[K_4,K_3,K_2,K_1,K_5;m_3,m_2,m_1,M_2,M_1]& = &  \WH g_5^{-3}\WH g_2 {\WH I}_{5}\nn
I_5(n_1,n_5,n_4,n_3,n_2)[K_5,K_4,K_3,K_2,K_1;M_1,m_3,m_2,m_1,M_2]& = &  \WH g_5^{-4}\WH g_2 {\WH I}_{5}~~~\label{Pen-sym-2}\eea
Furthermore, using the same idea in the subsection \ref{Tri-recur} we can write down
similar recurrence relation for general $I_5(n_1,n_2,n_3,n_4,n_5)$ using the expansion
of $I_5(2,1,1,1,1)$, $I_5(1,2,1,1,1)$, $I_5(1,1,2,1,1)$, $I_5(1,1,1,2,1)$ and $I_5(1,1,1,1,2)$. However,
by relation \eref{Pen-sym-1} and \eref{Pen-sym-2}, all other four cases  can be reduced to the reduction of   $I_5(1,1,1,1,2)$.

To reduce $I_5(1,1,1,1,2)$, according to our ideas, we should  calculate $\frac{\d}{\d m_3^2}I_5(1,1,1,1,1)$ in $\binom{5}{2}=10$ different cuts. Again, in the main part, we present
only the computation of the cut $K_1$.
For this case, since  the analytic checking using LiteRed \cite{Lee:2013mka} is too hard, we have
checked only numerically.

%%%%%%%%%%%%%%%%%%%%%%%%%%%%%%%%%
\subsection{Cut $K_1$ of $I_2(1,1,1,1,2)$}
%%%%%%%%%%%%%%%%%%%%%%%%%%%%%%%%%%

For this cut, we will choose the parameters of the second form in  \eref{Pen-gen-1-1}
as $K=K_1$, $P_1=K_{12}$, $P_2=K_{123}$ and $P_3=-K_5$.
Since the $m_3$ is not contained in $(\frac{\Delta}{4K^2})^{-\eps}$, we could just drop the factor $(\frac{\Delta}{4K^2})^{-\eps}$.
Furthermore the parameter $m_3$ is only contained in $B$, $D$, $H_0$ and $T_{12}$, $T_{13}$ and $T_{23}$, thus we have
\bea
\frac{\d}{\d m_3^2}C(I_5)&=&\frac{b}{4(K^2)^2}\Big(\frac{\d T_{23}}{\d m_3^2}pen_{23}^{(0)}+\frac{\d T_{13}}{\d m_3^2}pen_{13}^{(0)}+\frac{\d T_{12}}{\d m_3^2}pen_{12}^{(0)}\nn
&&+T_{23}\frac{\d}{\d m_3^2}pen_{23}^{(0)}+T_{13}\frac{\d}{\d m_3^2}pen_{13}^{(0)}+T_{12}\frac{\d}{\d m_3^2}pen_{12}^{(0)}\Big)~~\label{116}
\eea
The result of $\frac{\d T_{ij}}{\d m_3^2}$ is very simple and given by
\bea
\frac{\d T_{23}}{\d m_3^2}=8(\gamma_{12}\gamma_{23}-\beta_{2}\gamma_{13}),~~~
\frac{\d T_{13}}{\d m_3^2}=8(\gamma_{12}\gamma_{13}-\beta_1\gamma_{23}),~~~
\frac{\d T_{12}}{\d m_3^2}=8(\beta_1\beta_2-\gamma_{12}^2)
\eea
To calculate $\frac{\d}{\d m_3^2}pen_{ij}^{(0)}$, noticing that
\bea
\frac{\d}{\d m_3^2}pen^{(0)}&=&\int_0^1duu^{-1-\eps}\frac{1}{(H_0+H_1u)^2}\frac{-\frac{\d H_0}{\d m_3^2}}{\sqrt{B-Au}}\ln(*)
+\int_0^1duu^{-1-\eps}\frac{1}{H_0+H_1u}\frac{-\frac{1}{2}\frac{\d B}{\d m_3^2}}{\sqrt{B-Au}^3}\ln(*)\nn
&&+\int_0^1duu^{-1-\eps}\frac{1}{H_0+H_1u}\frac{1}{\sqrt{B-Au}}\frac{\d}{\d m_3^2}\ln(*)
\equiv L_1+L_2+L_3~~~~\label{L123}
\eea
 we need to calculate  these three terms respectively.

To prepare the reduction of  three integral $L_1$, $L_2$, and $L_3$, we rewrite  $pen^{(n)}$, which is defined in \eref{Pen-def-1-6}, as following
To prepare the reduction of  three integral $L_1$, $L_2$, and $L_3$, we want to rewrite  $pen^{(n)}$ defined in \eref{Pen-def-1-6} in the following by doing the partial integration $du^{n-\eps}=(n-\eps)u^{n-1-\eps}du$ 
\bea
pen^{(n)}&=&\frac{1}{n-\eps}\Big\{H_{1}\int_0^1duu^{n-\eps}\frac{1}{(H_0+H_{1}u)^2}\frac{1}{\sqrt{B-Au}}\ln(*)\nn
&&-\frac{A}{2}\int_0^1duu^{n-\eps}\frac{1}{H_0+H_{1}u}\frac{1}{\sqrt{B-Au}^3}\ln(*)\nn
&&+\int_0^1duu^{n-\eps}\frac{AD+BD-2BC+u(AC+BC-2AD)}{(H_0+H_{1}u)(B-Au)}\frac{1}
{\sqrt{1-u}~\$}\Big\}~~\label{pen-ibp}
\eea
where
\bea
\$_{ij}&=&(D_{ij}-C_{ij}u)^2-(1-u)(B_{ij}-A_{ij}u)=\beta_i\beta_j(1-u-Z_i^2)(1-u-Z_j^2)
~~~~\label{dollar-ij}
\eea
The first line in \eref{pen-ibp} has the form of  $L_1$ in \eref{L123} since $H_0$ does not depend on $u$ as defined in \eref{Pen-def-1-2}.

Now we consider the third line in \eref{pen-ibp}. Using algebraic separation
%%
%\bea
%\frac{1}{(H_0+H_{1}u)(B-Au)}&=&\frac{H_1}{AH_0+BH_1}\frac{1}{H_0+H_{1}u}+\frac{A}{AH_0+BH_1}\frac{1}{B-Au}\nn
%\frac{u}{(H_0+H_{1}u)(B-Au)}&=&\frac{-H_0}{AH_0+BH_{1}}\frac{1}{H_0+H_{1}u}+\frac{B}{AH_0+BH_{1}}\frac{1}{B-Au}
%\eea
%%
it becomes
\bea
&&\int_0^1duu^{n-\eps}\frac{AD+BD-2BC+u(AC+BC-2AD)}{(H_0+H_{1}u)(B-Au)}\frac{1}{\sqrt{1-u}\$}\nn
&=&\frac{(AD+BD-2BC)H_1-H_0(AC+BC-2AD)}{AH_0+BH_1}\int_0^1duu^{n-\eps}\frac{1}{(H_0+H_1u)\sqrt{1-u}\$}\nn
&&+\frac{(AD+BD-2BC)A+B(AC+BC-2AD)}{AH_0+BH_1}\int_0^1duu^{n-\eps}\frac{1}{(B-Au)\sqrt{1-u}\$}
~~\label{sp5}
\eea
Among these two terms in \eref{sp5}, the second term will be canceled by the second line in \eref{pen-ibp}. For the first term,
 using the factorization form of $\$_{ij}$ in \eref{dollar-ij}
%and  the spliting relation
%
%\bea
%\frac{1}{(1-u-Z_i^2)(1-u-Z_j^2)}&=&\frac{1}{Z_i^2-Z_j^2}(\frac{1}{1-u-Z_i^2}-\frac{1}{1-u-Z_j^2})\nn
%\frac{u}{(1-u-Z_i^2)(1-u-Z_j^2)}&=&\frac{1}{Z_i^2-Z_j^2}(\frac{1-Z_i^2}{1-u-Z_i^2}
%-\frac{1-Z_2^2}{1-u-Z_2^2})~~~\label{spi-u}
%\eea
%
we get
\bea
& & \frac{1}{(H_0+H_1u)\sqrt{1-u}\$_{ij}}=\frac{1}{\sqrt{1-u}\beta_i\beta_j(Z_i^2-Z_j^2)}\times\frac{1}{H_0+H_1u}(\frac{1}{1-u-Z_i^2}-\frac{1}{1-u-Z_j^2})\nn
& =& W_{0;ij}\frac{1}{\sqrt{1-u}(H_0+H_1u)} +W_{i;ij}\frac{1}{\sqrt{1-u}(1-u-Z_i^2)}+W_{j;ij}\frac{1}{\sqrt{1-u}(1-u-Z_j^2)}~~~\label{h0h1}
\eea
with the coefficients
\bea
W_{0;ij}&=&\frac{1}{\beta_i\beta_j(Z_i^2-Z_j^2)}\Big(\frac{H_1}{H_0+(1-Z_i^2)H_1}-\frac{H_1}{H_0+(1-Z_j^2)H_1}\Big)\nn
W_{i;ij}&=&\frac{1}{\beta_i\beta_j(Z_i^2-Z_j^2)}\frac{1}{H_0+(1-Z_i^2)H_1},~~~
W_{j;ij}=\frac{1}{\beta_i\beta_j(Z_i^2-Z_j^2)}\frac{-1}{H_0+(1-Z_j^2)H_1}~~~\label{Wij-def}
\eea
Among three terms in \eref{h0h1}, the first term is spurious and will be canceled when summing
contributions from three $pen_{ij}$ terms. The second and third terms are essentially the triangle part. Putting all coefficients back, we find
the third line in \eref{pen-ibp} is given by
\bea
&&\int_0^1duu^{n-\eps}\frac{AD+BD-2BC+u(AC+BC-2AD)}{(H_0+H_1u)(B-Au)}\frac{1}{\sqrt{1-u}\$}\nn
&=&\frac{(A-B)(AD-BC)}{AH_0+BH_1}\int_0^1duu^{n-\eps}\frac{1}{(B-Au)\sqrt{1-u}\$}
+G_{ij}W_{0;ij}\int_0^1duu^{n-\eps}\frac{1}{\sqrt{1-u}(H_0+H_1u)}\nn
&&-\frac{G_{ij}W_{i;ij}(n-\eps)}{Z_i}Tri^{(n)}(Z_i)
-\frac{G_{ij}W_{j;ij}(n-\eps)}{Z_j}Tri^{(n)}(Z_j)~~~~~\label{last-done}
\eea
where
\bea
G_{ij}\equiv \frac{(A_{ij}D_{ij}+B_{ij}D_{ij}-2B_{ij}C_{ij})H_1-H_0(A_{ij}C_{ij}+B_{ij}C_{ij}
-2A_{ij}D_{ij})}{(A_{ij}H_0+B_{ij}H_1)}~~~\label{Gij-def}
\eea

For the second line in \eref{pen-ibp}, we will use  the rewriting of box
\bea
Box^{(n)}&=&\frac{1}{n-\eps}\times\Big\{-\frac{A}{2}\int_0^1duu^{n-\eps}\frac{1}{\sqrt{B-Au}^3}\ln(*)\nn
&&+\int_0^1duu^{n-\eps}\frac{AD+BD-2BC+u(AC+BC-2AD)}{(B-Au)\sqrt{1-u}\$}\Big\}~~\label{box-ibp}
\eea
which is obtained by the similar method as in \eref{pen-ibp}. Using the similar splitting technique done in previous paragraphes
to the second term in \eref{box-ibp}, we will arrive
\bea
& & Box_{ij}^{(n)}=\frac{1}{n-\eps}\Big\{\frac{-A}{2}\int_0^1duu^{n-\eps}\frac{1}{\sqrt{B-Au}^3}\ln(*)
+\frac{(A-B)(AD-BC)}{A}\int_0^1duu^{n-\eps}\frac{1}{(B-Au)\sqrt{1-u}\$}\nn
&&-\frac{2AD-AC-BC}{A}\frac{1}{\beta_i\beta_j(Z_i^2-Z_j^2)}\frac{n-\eps}{Z_i}Tri^{(n)}(Z_i)
+\frac{2AD-AC-BC}{A}\frac{1}{\beta_i\beta_j(Z_i^2-Z_j^2)}\frac{n-\eps}{Z_j}Tri^{(n)}(Z_j)\Big\}
~~~~~~\label{box1}
\eea
For the first term in \eref{box1}, we do following manipulation
\bea
&&\int_0^1duu^{n-\eps}\frac{1}{\sqrt{B-Au}^3}\ln(*)=\int_0^1duu^{n-\eps}\frac{H_0+H_1u}{(H_0+H_1u)
\sqrt{B-Au}^3}\ln(*)=\int_0^1duu^{n-\eps}\frac{H_0+H_1\frac{Au-B+B}{A}}{(H_0+H_1u)\sqrt{B-Au}^3}\ln(*)\nn
&&=(H_0+\frac{BH_1}{A})\int_0^1duu^{n-\eps}\frac{1}{(H_0+H_1u)\sqrt{B-Au}^3}\ln(*)-\frac{H_1}{A}pen^{(n+1)}~~~\label{box2}
\eea
Thus the second line in \eref{pen-ibp} becomes
\bea
& & -\frac{A}{2}\int_0^1duu^{n-\eps}\frac{1}{(H_0+H_1u)\sqrt{B-Au}^3}\ln(*)
=\frac{A(n-\eps)}{AH_0+BH_1}Box^{(n)}-\frac{AH_1}{2(AH_0+BH_1)}pen^{(n+1)}\nn
&&-\frac{(A-B)(AD-BC)}{(AH_0+BH_1)}\int_0^1duu^{n-\eps}\frac{1}{(B-Au)\sqrt{1-u}\$}
-\frac{A\lambda_{i;ij}(n-\eps)}{2}Tri^{(n)}(Z_i)-\frac{A\lambda_{j;ij}(n-\eps)}{2}Tri^{(n)}(Z_j)
~~~~~~~~\label{second-done}
\eea
 with the coefficients
\bea
\lambda_{i;ij}&=&\frac{2(AC+BC-2AD)}{A(AH_0+BH_1)}\frac{1}{\beta_i\beta_j(Z_i^2-Z_j^2)}
\frac{1}{Z_i},~~~
\lambda_{j;ij}=\frac{2(AC+BC-2AD)}{A(AH_0+BH_1)}\frac{1}{\beta_i\beta_j(Z_j^2-Z_i^2)}
\frac{1}{Z_j}~~~~\label{laij-def}
\eea
Now we can explicitly see that the third term in \eref{second-done} cancel the first term in
\eref{last-done}.

Putting all together, we have
\bea
& & pen^{(n)}=\frac{H_1}{n-\eps}\int_0^1duu^{n-\eps}\frac{1}{(H_0+H_1u)^2}\frac{1}{\sqrt{B-Au}}
\ln(*)+\frac{A}{AH_0+BH_1}Box^{(n)}-\frac{AH_1}{2(AH_0+BH_1)}\frac{1}{n-\eps}pen^{(n+1)}\nn
&&-\Big(\frac{A\lambda_{i;ij}}{2}+\frac{G_{ij}W_{i;ij}}{Z_i}\Big)Tri^{(n)}(Z_i)
-\Big(\frac{A\lambda_{j;ij}}{2}+\frac{G_{ij}W_{j;ij}}{Z_j}\Big)Tri^{(n)}(Z_j)
+\frac{G_{ij}W_{0;ij}}{n-\eps}\int_0^1duu^{n-\eps}\frac{1}{\sqrt{1-u}(H_0+H_1u)}~~~~~~~~\label{penn-1}
\eea
with the coefficients given by  \eref{Wij-def}, \eref{Gij-def} and \eref{laij-def}.

Now we can consider the $L_1, L_2, L_3$ terms. Let us start from the $L_1$ term. By comparing $L_1$ with the first term in \eref{L123},  we see that
\bea
L_1&=&-\frac{\d H_0}{\d m_3^2}\int_0^1duu^{-1-\eps}\frac{1}{(H_0+H_1u)^2\sqrt{B-Au}}\ln(*)\nn
&=&\frac{\d H_0}{\d m_3^2}\frac{(1+\eps)}{H_1}pen^{(-1)}-\frac{\d H_0}{\d m_3^2}\frac{A(1+\eps)}{H_1(AH_0+BH_1)}Box^{(-1)}-\frac{\d H_0}{\d m_3^2}\frac{A}{2(AH_0+BH_1)}pen^{(0)}\nn
&&+\frac{\d H_0}{\d m_3^2}(\frac{A\lambda_{i;ij}}{2}+\frac{G_{ij}W_{i;ij}}{Z_i})\frac{(1+\eps)}{H_1}Tri^{(-1)}(Z_i)+\frac{\d H_0}{\d m_3^2}(\frac{A\lambda_{j;ij}}{2}+\frac{G_{ij}W_{j;ij}}{Z_j})\frac{(1+\eps)}{H_1}Tri^{(-1)}(Z_j)\nn
&&+\frac{\d H_0}{\d m_3^2}\frac{G_{ij}W_{0;ij}}{H_1}\int_0^1duu^{-1-\eps}\frac{1}{\sqrt{1-u}(H_0+H_1u)}~~~~\label{l1part}
\eea
For the $L_2$, which is written as
\bea
L_2&=&-\frac{1}{2}\frac{\d B}{\d m_3^2}\int_0^1duu^{-1-\eps}\frac{1}{(H_0+H_1u)\sqrt{B-Au}^3}\ln(*)
\eea
using the \eref{box2} with $n=-1$ we have
\bea
L_2&=&-\frac{\d B}{\d m_3^2}\frac{(1+\eps)}{AH_0+BH_1}Box^{(-1)}-\frac{1}{2}\frac{\d B}{\d m_3^2}\frac{H_1}{AH_0+BH_1}pen^{(0)}\nn
&&-\frac{\d B}{\d m_3^2}\frac{(A-B)(AD-BC)}{A(AH_0+BH_1)}\int_0^1duu^{-1-\eps}\frac{1}{(B-Au)\sqrt{1-u}\$}\nn
&&+\frac{1}{2}\frac{\d B}{\d m_3^2}\lambda_{i;ij}(1+\eps)Tri^{(-1)}(Z_i)+\frac{1}{2}\frac{\d B}{\d m_3^2}\lambda_{j;ij}(1+\eps)Tri^{(-1)}(Z_j)~~~~\label{L2-result}
\eea
For the integral $L_3$
\bea
L_3
&=&\int_0^1duu^{-1-\eps}\frac{1}{(H_0+H_1u)}\frac{(1-u)[B'D-2BD'+u(2AD'-B'C)]}{(B-Au)\sqrt{1-u}\$}
\eea
where $B'\equiv\frac{\d B}{\d m_3^2}$, $D'\equiv\frac{\d D}{\d m_3^2}$, using the splitting
of $\S$ in \eref{dollar-ij} it is given by
\bea
L_3&=&r_1\int_0^1duu^{-1-\eps}\frac{1}{\sqrt{1-u}\$}
+r_2\int_0^1duu^{-1-\eps}\frac{1}{(B-Au)\sqrt{1-u}\$}
+r_3\int_0^1duu^{-1-\eps}\frac{1}{(H_0+H_1u)\sqrt{1-u}\$}~~~~~~~
\eea
with the coefficients
\bea
r_{1}&=&\frac{2AD'-B'C}{AH_1},~~~
r_{2}=\frac{B'(A-B)(AD-BC)}{A(AH_0+BH_1)}\nn
r_{3}&=&\frac{(H_0+H_1)\Big[(B'D-2BD')H_1-(2AD'-B'C)H_0\Big]}{H_1(AH_0+BH_1)}
\eea
There are three terms. The first term could be split into two triangles, as we have done before. The second term is a spurious term, and is canceled with the same term in the integral $L_2$. And the last term could also be split into three pieces by using \eref{last-done} with $n=-1$. Putting all together we have
\bea
L_3&=&r_2\int_0^1duu^{-1-\eps}\frac{1}{(B-Au)\sqrt{1-u}\$}
+r_{3}W_{0;ij}\int_0^1duu^{-1-\eps}\frac{1}{\sqrt{1-u}(H_0+H_1u)}\nn
&&+\Big(\frac{r_1}{\beta_i\beta_j(Z_i^2-Z_j^2)}+r_3W_{i;ij}\Big)\frac{1+\eps}{Z_i}Tri^{(-1)}(Z_i)
+\Big(\frac{r_1}{\beta_i\beta_j(Z_j^2-Z_i^2)}+r_{3}W_{j;ij}\Big)\frac{1+\eps}{Z_j}Tri^{(-1)}(Z_j)
~~~~~
\eea

Collecting above results for $L_1, L_2, L_3$ we have
\bea
\frac{\d }{\d m_3^2}pen_{ij}^{(0)}&=&\frac{H_0'(1+\eps)}{H_1}pen^{(-1)}-\frac{(AH_0'+H_1B')}{2(AH_0+BH_1)}pen^{(0)}
+\frac{(1+\eps)}{AH_0+BH_1}\Big(-\frac{AH_0'}{H_1}-B'\Big)Box^{(-1)}
\nn
& & +\frac{(1+\eps)}{2}\Big(\frac{2r_{3}W_{i;ij}}{Z_i}+\frac{2r_{1}}{\beta_i\beta_jZ_i(Z_i^2-Z_j^2)}+B'\lambda_{i;ij}+\frac{H_0'(2G_{ij}W_{i;ij}+AZ_i\lambda_{i;ij})}{H_1Z_i}\Big)Tri^{(-1)}(Z_i)\nn
&&+\frac{(1+\eps)}{2}\Big(\frac{2r_{3}W_{j;ij}}{Z_j}+\frac{2r_1}{\beta_i\beta_jZ_j(Z_j^2-Z_i^2)}+B'\lambda_{j;ij}+\frac{H_0'(2G_{ij}W_{j;ij}+AZ_j\lambda_{j;ij})}{H_1Z_j}\Big)Tri^{(-1)}(Z_j)\nn
&&+\Big(\frac{H_0'G_{ij}W_{0;ij}}{H_1}+r_{3}W_{0;ij}\Big)\int_0^1duu^{-1-\eps}\frac{1}
{\sqrt{1-u}(H_0+H_1u)}~~~\label{result-pre}
\eea
In \eref{result-pre}, the last term will be canceled when summing over three $\frac{\d }{\d m_3^2}pen_{ij}^{(0)}$ in \eref{116}.
To continue, we need to reduce $Tri^{(-1)}$, $Box^{(-1)}$ and $pen^{(-1)}$ to our scalar basis.
Using
\bea
H_0pen^{(n)}+H_1pen^{(n+1)}&=&Box^{(n)}
\eea
with $n=-1$,
\bea
Tri^{(-1)}(Z)=\frac{\eps}{(\eps+1)(1-Z^2)}Tri^{(0)}(Z)+\frac{(1-2\eps)Z}{(1-Z^2)(\eps+1)}Bub^{(0)}
\eea
and
\bea
Box^{(-1)}&=&\frac{\frac{1}{2}+\eps}{1+\eps}\frac{A}{B}Box^{(0)}+\frac{\eps}{1+\eps}\frac{C_{Z_1}}{BZ_1(1-Z_1^2)}Tri^{(0)}(Z_1)+\frac{\eps}{1+\eps}\frac{C_{Z_2}}{BZ_2(1-Z_2^2)}Tri^{(0)}(Z_2)\nn
& &+\frac{1-2\eps}{1+\eps}\Big(\frac{C_{Z_1}}{(1-Z_1^2)B}+\frac{C_{Z_2}}{(1-Z_2^2)B}\Big)Bub^{(0)}
\eea
we have
\bea
\frac{\d}{\d m_3^2}pen_{ij}^{(0)}&=&q_{ij;5}pen_{ij}^{(0)}+q_{ij;4}Box_{ij}^{(0)}+q_{ij;Z_i}Tri^{(0)}(Z_i)+q_{ij;Z_j}Tri^{(0)}(Z_j)+q_{ij;2}Bub^{(0)}~~\label{bbb}
\eea
with the coefficients
\bea
q_{ij;5}&=&l_{ij;5}=-\frac{H_0'(1+\eps)}{H_0}-\frac{(A_{ij}H_0'+H_1B_{ij}')}{2(A_{ij}H_0+B_{ij}H_1)}\nn
q_{ij;4}&=&\frac{\frac{1}{2}+\eps}{1+\eps}\frac{A}{B}l_{ij;4}=\frac{A_{ij}(B_{ij}H_0'-B_{ij}'H_0)(\frac{1}{2}+\eps)}{B_{ij}H_0(A_{ij}H_0+B_{ij}H_1)}\nn
q_{ij;Z_i}&=&l_{ij;4}\times \frac{\eps}{1+\eps}\frac{C_{Z_i;ij}}{B_{ij}Z_i(1-Z_i^2)}+l_{ij;Z_i}\times\frac{\eps}{(\eps+1)(1-Z_i^2)}\nn
&=&\frac{q_{i;ij;n}}{B_{ij}H_0(A_{ij}H_0+B_{ij}H_1)Z_i(Z_i^2-1)(H_0+H_1(1-Z_i^2))(Z_i^2-Z_j^2)\beta_i\beta_j}\nn
q_{ij;Z_j}&=&l_{ij;4}\times\frac{\eps}{1+\eps}\frac{C_{Z_j;ij}}{B_{ij}Z_j(1-Z_j^2)}+l_{ij;Z_2}\times\frac{\eps}{(\eps+1)(1-Z_j^2)}\nn
&=&\frac{q_{j;ij;n}}{B_{ij}H_0(A_{ij}H_0+B_{ij}H_1)Z_j(Z_j^2-1)(H_0+H_1(1-Z_j^2))(Z_j^2-Z_i^2)\beta_i\beta_j}\nn
q_{ij;2}&=&l_{ij;4}\times\frac{1-2\eps}{1+\eps}\Big(\frac{C_{Z_i;ij}}{(1-Z_i^2)B_{ij}}+\frac{C_{Z_j;ij}}{(1-Z_j^2)B_{ij}}\Big)\nn
&&+l_{ij;Z_i}\times\frac{(1-2\eps)Z_i}{(1-Z_i^2)(\eps+1)}+l_{ij;Z_j}\times\frac{(1-2\eps)Z_j}{(1-Z_j^2)(\eps+1)}~~~\label{ccc}
\eea
where
\bea
q_{i;ij;n}&=&\eps\Big[B^2H_0(-DH_0'+2D'H_1Z_i^2+CH_0'(1+Z_i^2))+B'C_{Z_{i}}H_0(H_0+H_1(1-Z_i^2))(Z_i^2-Z_j^2)\beta_i\beta_j\nn
&&+B\Big(AH_0(DH_0'+(2D'H_0-2DH_0')Z_i^2+CH_0(Z_i^2-1))+B'H_0(D(H_0+H_1(1-2Z_i^2))\nn
& & -C(H_0+H_1+H_1Z_i^2-H_1Z_i^2))-C_{Z_i}H_0'(H_0+H_1-H_1Z_i^2)(Z_i^2-Z_j^2)\beta_i\beta_j\Big)\Big]\nn
q_{j;ij;n}&=&q_{j;ij}|_{i\leftrightarrow j},~~~B'\equiv \frac{\d B}{\d m_3^2},~~~
D'\equiv \frac{\d D}{\d m_3^2},~~~
H_0'\equiv \frac{\d H_0}{\d m_3^2},~~~
C_{Z_k;ij}=D_{ij}+(Z_k^2-1)C_{ij}
\eea

Now we  put the result \eref{bbb} to \eref{116} to get
\bea
\frac{\d}{\d m_3^2}C(I_5)&=&\frac{b}{4(K^2)^2}\Big\{(\frac{\d T_{12}}{\d m_3^2}+q_{12;5}T_{12})pen_{12}^{(0)}+(\frac{\d T_{13}}{\d m_3^2}+q_{13;5}T_{13})pen_{13}^{(0)}+(\frac{\d T_{23}}{\d m_3^2}+q_{23;5}T_{23})pen_{23}^{(0)}\nn
&&+q_{12;4}T_{12}Box_{12}^{(0)}+q_{13;4}T_{13}Box_{13}^{(0)}+q_{23;4}T_{23}Box_{23}^{(0)}+(q_{12;Z_1}T_{12}+q_{13;Z_1}T_{13})Tri^{(0)}(Z_1)\nn
&&+(q_{12;Z_2}T_{12}+q_{23;Z_2}T_{23})Tri^{(0)}(Z_2)+(q_{13:Z_3}T_{13}+q_{23;Z_3}T_{23})Tri^{(0)}(Z_3)\nn
&&+(q_{12;2}T_{12}+q_{13;2}T_{13}+q_{23;2}T_{23})Bub^{(0)}\Big\}~~~\label{K1-m3}
\eea
where $K=K_1$
and the parameters are given in \eref{ccc} and  \eref{para}. One important point is that
the sum of the first three terms in \eref{K1-m3} gives exactly the cut of pentagon in \eref{ci5}.
Thus  we have the result
\bea
I_5(1,1,1,1,2)&=&c_{5\to5;K_1}I_5+c_{5\to 4;\bar5;K_1}I_{4;\bar5}+c_{5\to3;\bar4;K_1}I_{4;\bar4}+c_{5\to4;\bar3;K_1}I_{4;\bar3}+c_{5\to 3;\bar4\bar5;K_1}I_{3;\bar4\bar5}+c_{5\to3;\bar3\bar5;K_1}I_{3;\bar3\bar5}
\nn & & +c_{5\to3;\bar3\bar4;K_1}I_{3;\bar3\bar4}+c_{5\to2;\bar3\bar4\bar5;K_1}I_{2;\bar3\bar4\bar5}+\cdots
\eea
where the elips represents the tadpoles and the coefficients are given by
\bea
c_{5\to5;K_1}&=&\frac{1}{T_{12}}\frac{\d T_{12}}{\d m_3^2}+q_{12}=\frac{1}{T_{13}}\frac{\d T_{13}}{\d m_3^2}+q_{13}=\frac{1}{T_{23}}\frac{\d T_{23}}{\d m_3^2}+q_{23}\nn
c_{5\to4;\bar3;K_1}&=&\frac{1}{2K^2}q_{23;4}T_{23},~~~
c_{5\to4;\bar4}=\frac{1}{2K^2}q_{13;4}T_{13},~~~
c_{5\to4\bar5}=\frac{1}{2K^2}q_{12;4}T_{12}\nn
c_{5\to3;\bar4\bar5;K_1}&=&\frac{-b\sqrt{\Delta_{3;m=0[Z_1]}}}
{4K^4}(q_{12;Z_1}T_{12}+q_{13;Z_1}T_{13}),~~~\Delta_{3;m=0}[Z_1]=4[(K\cdot P_1)^2-K^2P_1^2]\nn
c_{5\to3;\bar3\bar5;K_1}&=&\frac{-b\sqrt{\Delta_{3;m=0[Z_2]}}}{4K^4}(q_{12;Z_2}
T_{12}+q_{23;Z_2}T_{23}),~~~\Delta_{3;m=0}[Z_2]=4[(K\cdot P_2)^2-K^2P_2^2]\nn
c_{5\to3;\bar3\bar4;K_1}&=&\frac{-b\sqrt{\Delta_{3;m=0[Z_3]}}}{4K^4}(q_{13;Z_3}T_{13}
+q_{23;Z_3}T_{23}),~~~\Delta_{3;m=0}[Z_3]=4[(K\cdot P_3)^2-K^2P_3^2]\nn
c_{5\to2;\bar3\bar4\bar5;K_1}&=&\frac{1}{4K^4}(q_{12;2}T_{12}+q_{13;2}T_{13}+q_{23;2}T_{23}),~~~
P_1=K_{12},~~~
P_2=K_{123},~~~
P_3=K_{1234}~~~~\label{K1-c}
\eea
with the parameters given in \eref{ccc}. The corresponding basis are
\bea
&&I_{4;\bar5}=I_5(1,1,1,1,0),~~~I_{4;\bar4}=I_5(1,1,1,0,1),~~~I_{4;\bar3}=I_5(1,1,0,1,1)\nn
&&I_{3;\bar4\bar5}=I_5(1,1,1,0,0),~~~I_{3;\bar3\bar5}=I_5(1,1,0,1,0),~~~I_{3;\bar3\bar4}=I_5(1,1,0,0,1),~~~I_{2;\bar3\bar4\bar5}=I_5(1,1,0,0,0)~~~~~~~~~~~~~\label{Basis-K1}
\eea

Again, having presented  details for the cut $K_1$, the computation of other cuts will be similar, as shown in the section of triangle. For simplicity we will not list them one by one. All coefficients have been checked using LiteRed \cite{Lee:2013mka} numerically.

%%%%%%%%%%%%%%%%%%%%%%
\section{Conclusion}
%%%%%%%%%%%%%%%%%%%%%%

In this paper, we have considered the reduction of one-loop integrals with higher poles using the unitarity cut method.
By   the trick of differentiation over auxiliary masses, we have translate the problem to the
decomposition of differentiation of imaginary part of scalar basis. Furthermore, from the angle of
differentiation, recurrence relation can be established and the whole reduction can be
reduced to the reduction of basic integral, i.e., one and only one propagator with power two.
 We demonstrate our method by
carrying out the reduction of basic example for the scalar bubbles, triangles, boxes and pentagons and give
analytic expression for reduction coefficients.

One of unsolved problems of our algorithm is the analytic tadpole coefficients of the reduction. Although the recurrence relation can be established for the tadpole coefficients, the tadpole
coefficients for the basic integrals, for examples $I_2(1,2)$, $I_3(1,1,2)$, $I_4(1,1,1,2)$ and
$I_5(1,1,1,1,2)$ can not be found by unitarity cut method. One can use the technique developed in \cite{Britto:2009wz,Britto:2010um,Britto:2012mm} or the familiar IBP method to calculate them to complete the whole reduction program for higher poles.

%
%%%%%%%%%%%%%%%%%%%%%%%%%%%%%%%%%%%%%%%%%%
%\newpage
%%%%%%%%%%%%%%%%%%%%%%%%%%%%%%%%%%%%%%%%%%%
\section*{Acknowledgments}
%%%%%%%%%%%%%%%%%%%%%%%%%%%%%%%%%%%%%%%%%

We would like to thank Y.Zhang for very useful discussion and reading of
draft. This work is supported by Qiu-Shi Funding and Chinese NSF
funding under Grant  No.11935013, No.11947301, No.12047502 (Peng Huanwu Center).

%%%%%%%%%%%%%%%%%%%%%%%%%%%%%%%%%%%%%%%%%%%%%%%%%%%%%%%%%%%%%%%%%%%%

%\newpage
\bibliographystyle{JHEP}

\bibliography{reference1}

%\bibitem{Snirnov}
%V.A. Smirnov, "Evaluating Feynman Integrals", Springer

%%%%%%%%%%%%%%%%%%%%%%%%%%%%%%%%%%%%%%%%%%%%%%%%%%%%%%%%%
\end{document}